\documentclass[preprint,
 reprint,
 onecolumn,
superscriptaddress,
amsmath,amssymb,
 aps,
]{revtex4-2}

\usepackage{graphicx}
\usepackage[utf8]{inputenc}
\usepackage[T1]{fontenc}
\usepackage{tabularx}
\usepackage{xcolor}
\usepackage{bm}
\usepackage{soul}
\usepackage{xr}
\usepackage{float}

\usepackage{graphicx}  
\usepackage{microtype} 
\usepackage{wasysym}

\begin{document}

\preprint{APS/123-QED}

\title{Dramatic pressure-induced slowdown and \\ the soft glassy rheology of water at 2D material interface}

\author{Guilhem Mariette}
\affiliation{Soft Matter Sciences and Engineering, ESPCI Paris, PSL University, CNRS, Sorbonne Université, 75005 Paris, France}%
\author{Kenji Watanabe}
\affiliation{Research Center for Electronic and Optical Materials, National Institute for Materials Science, 1-1 Namiki, Tsukuba 305-0044, Japan}

\author{Takashi Taniguchi}
\affiliation{Research Center for Materials Nanoarchitectonics, National Institute for Materials Science,  1-1 Namiki, Tsukuba 305-0044, Japan}

\author{Jean Comtet}%
 \email{jean.comtet@espci.fr}
\affiliation{Soft Matter Sciences and Engineering, ESPCI Paris, PSL University, CNRS, Sorbonne Université, 75005 Paris, France}%

\date{\today}

\begin{abstract}\textbf{
The flow of molecularly confined water has attracted considerable interest in recent years. Yet despite their exquisite character, current nanofluidic devices cannot probe molecular water flow beyond the linear-response regime, while nanomechanical force-based approaches have produced markedly inconsistent results. Here, using tuning-fork-based atomic force microscopy, we tackle these limitations by measuring the full linear-to-nonlinear, shear-dependent viscoelastic rheology of molecularly confined water at various two-dimensional material interfaces -- demonstrating pronounced rate effects in water flow response. The observed rheological behaviors are remarkably analogous to that of macroscopic soft glassy yielding materials and fully captured by a non-linear model accounting for the shear-induced acceleration of the dynamics -- a hallmark of disordered supercooled and metastable systems. We further unveil a pronounced pressure-induced slowdown in nanoconfined water dynamics, with relaxation times extending into the microsecond range and marked material dependence. These results reveal non-linear, glassy rheology as a general feature of interfacial water and offer a route to tunable, non-linear nanofluidic elements.
}
\end{abstract}

\maketitle

\section*{Introduction}

The structure and dynamics of water at interfaces and under molecular-scale confinement is central to fields ranging from biology and electrochemistry to tribology. Over the past decade, nanofluidic devices have transformed this field, offering exquisite control over confinement geometry -- from the nanometric scale down to a single molecular layer -- and over the chemistry of the confining walls -- through well-defined 2D materials.
These experimental progress have revealed striking phenomena, including anomalous dielectric screening \cite{wang2025plane,fumagalli2018anomalously} and strongly non-linear ionic transport \cite{marcotte2020mechanically,mouterde2019molecular,cui2025coupling}. A parallel effort has targeted water flow itself \cite{radha2016molecular,keerthi2021water,secchi2016massive,allemand2023anomalous}, revealing a strong dependence on the chemical nature, but also curvature, of the confining material, with ultrafast flow through graphene-based 2D channels and nanotubes.
Yet, despite the immense sophistication of these devices, existing measurements are either restricted to large confinement of tens of nanometers \cite{secchi2016massive}, or rely on ensemble-averaged flow through parallelized molecular channels \cite{radha2016molecular,keerthi2021water,xie2018fast} subjected to poorly controlled capillary-driven flows. 
Therefore, none of these approaches can resolve the exact molecular flow profile at the interface \cite{schlaich2017hydration,carlson2025subnanometer}, nor go beyond a linear-response picture for water transport. 

A complementary route to probe interfacial and confined water dynamics comes from nanomechanical, force-based techniques -- Atomic Force Microscopy (AFM) and Surface Force Apparatus (SFA) -- which can probe the effective liquid rheology in a continuous manner from large distances down to molecular confinement \cite{lizee2024liquid}. Yet these approaches have produced starkly contradictory results for nanometric and molecularly confined flow: some report no more than a modest rise in viscosity \cite{raviv2004fluidity,raviv2001fluidity,raviv2002fluidity}; others, a dramatic rise upon nanoconfinement \cite{ortiz2013interplay,li2007structured,li2008nonlinear,zhu2001viscosity,goertz2007hydrophilicity,ma2015origins} spanning up to seven orders of magnitude \cite{cafolla2018lubricating}; and still others, a fully solid-like viscoelastic response \cite{jeffery2004direct,kim2013unified,li2014nanorheology,antognozzi2001observation}.

A natural resolution for these discrepancies would be that, as in macroscopic disordered and soft materials, the measured response depends on the excitation itself \cite{chen2010rheology} through the frequency $\omega$ or strain amplitude $\gamma$ of the applied perturbation, through an effective viscosity $\eta = G^{\prime\prime}(\omega,\gamma)/\omega$, with $G^{\prime\prime}$ the dissipative shear modulus. Such rate-dependent, non-linear rheology has indeed been established for confined organic \cite{bureau2010nonlinear,demirel1996glasslike} and ionic liquids \cite{bou2010nanoconfined}. For water, however, the evidence remains scarce and ambiguous \cite{khan2010dynamic,zhu2001viscosity,li2008nonlinear,kim2015probing}, hampered by the difficulty of isolating the linear from the non-linear flow regime. Reported signatures include a dependence on approach velocity \cite{khan2010dynamic,goertz2007hydrophilicity,khan2015squeeze}, on oscillation amplitude \cite{li2008nonlinear}, and on excitation frequency \cite{zhu2001viscosity} -- but no single study has cleanly separated the two regimes. This question merges with that of boundary lubrication, where mono- or bilayer-confined water (so-called 2D water) shows non-linear, typically logarithmic force-velocity relationships \cite{diao2019slippery,ma2015origins}. These have so far been captured only by solid-friction models based on thermally activated barrier hopping. Whether a genuine liquid-to-solid transition underlies this behavior, and how confined water responds to pressurized contact, thus remains unresolved.

Here, we address these questions by resolving the full linear-to-non-linear, shear-dependent response of water confined from nanometric scales down to two-dimensional, pressurized contact. Our measurements reveal key features reminiscent of the soft glassy rheology of yielding materials \cite{bonn2017yield,miyazaki2006nonlinear,hebraud1998mode,sollich1997rheology}, with a dissipation overshoot marking the yielding transition, followed by a power-law decay of the elastic and dissipative impedances at large oscillation amplitudes. Soft glassy rheology is typically associated with macroscopic disordered materials such as foams, emulsions, and granular systems, which are characterized by structural disorder and metastable character that give rise to flow behaviors reminiscent of glassy or supercooled liquids. These key features are thus expected to extend to confined pressurized water. This nanoscale rheological behavior is indeed fully captured by a non-linear rheological model accounting for the shear-induced acceleration of the dynamics -- a hallmark of supercooled systems. The relaxation time reaches the microsecond range and exhibits a strong increase with pressure,
 reflecting a continuous pressure-induced slowdown of water dynamics and molecular mobility. Finally, we uncover marked material dependence: while slowdown is most pronounced on hydrophilic mica, we find faster yet comparable water relaxation times on the hydrophobic HOPG and hBN. This feature is at odds with established results pointing to increased friction on the insulating hBN \cite{keerthi2021water, wu2024probing}, suggesting that interfacial relaxation and slip are governed by distinct mechanisms. 
 This non-linear "pasty" yielding response for confined water opens exciting prospects for relaxation-time engineering and the development of non-linear nanoscale fluidic elements. 
 
\section*{Results and Discussion}
\noindent \textbf{Experimental set-up and near-surface interactions.}

\begin{figure}[htb!]
    \centering
\includegraphics[width=1\textwidth, trim = 0cm 1.5cm 0cm 0cm, clip]{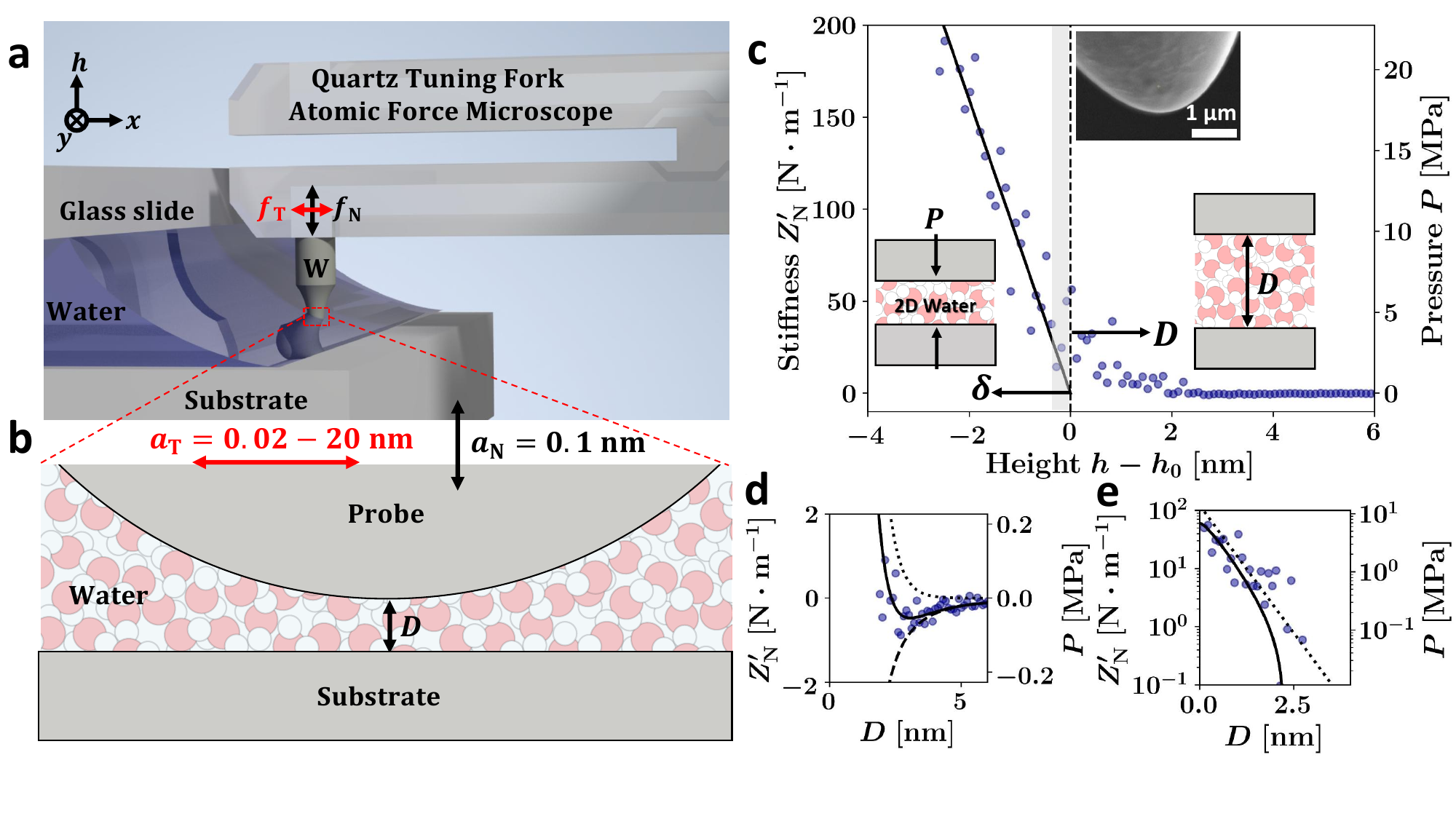}
\vspace{-10pt}
    \caption{\textbf{Quartz tuning fork–based atomic force microscope.} 
   \textbf{ (a)} A tungsten probe is glued to the quartz tuning fork (QTF) and immersed in water between a glass slide and a lamellar substrate. The QTF–probe system is excited by a piezo-dither in two modes: normal (tapping) and tangential (shear), relative to the planar substrate at $f_\text{N}\sim$32 kHz and $f_\text{T}\sim$18 kHz, with quality factors of $\sim$10,000 and $\sim$5,000, respectively. 
    \textbf{(b)} The probe records both normal (tapping) and shear responses of interfacial water on mica, HOPG, and hBN under nanometric confinement. The normal oscillation amplitude is kept constant at $a_\text{N}=100$ pm, while the tangential amplitude $a_\text{T}$ is varied between 20 pm and 20 nm. 
    \textbf{(c)} Load-stiffness curve for tungsten on hydrated HOPG. The black line is a linear fit, allowing the determination of the solid-contact position $h_0$ (emphasized by the dotted line). The indentation regime ($\delta =h_0 - h>0$) is dominated by a linear stiffness ascribed to the compliant oxide shell. The non-contact regime ($D=h - h_0>0$) is governed by van der Waals attraction for gap distance $D>3$ nm and by an intermediate repulsive interaction for $D\in[0,3]$ nm. The mechanical zero predicted by the van der Waals fit is indicated by the dashed line. The shaded gray region represents the 0.3~nm uncertainty associated with the two methods used to determine the solid contact position. Insets: Scanning electron microscopy image of a tungsten probe, showing a parabolic apex with a micrometric radius of curvature. Schematics showing (right) the non-contact region with confinement distance $D$ and (left) the contact regime where 2D water is confined at constant molecular distance with the load stiffness $Z^{\prime}_\text{N}$ controlling the applied pressure $P$ computed as $P = Z^{\prime}_\text{N}/2\pi R$.
   \textbf{ (d)} Magnified view of the transition from attraction to repulsion at $D \sim 3$ nm, fitted using the combined expression (solid line) of van der Waals attraction (dashed line) and exponential repulsion (dotted points). 
    \textbf{(e)} Log-lin view of the exponential-like repulsion dominating for $D\in[0,3]$ nm. }
\label{fig1:experimentalsetup}
\end{figure}
We start in Fig.~\ref{fig1:experimentalsetup} a--b with a description of the experimental setup. Briefly, we employ quartz tuning fork-based dynamic atomic force microscopy (AFM) to confine ultrapure water between tungsten probes of micrometric radius of curvature $R$ and various atomically flat lamellar 2D materials, namely mica, HOPG, and hBN. Glued to one prong of the tuning fork, the probe can be oscillated in both normal (tapping, black arrows) and tangential (shear, red arrows) motion with respect to the surface. Liquid confinement can be tuned from nanometric separation down to pressurized contact, where strongly bound hydration layers are compressed between the two solids, a situation we refer to as 2D confinement. By keeping the quartz crystal of the tuning fork out of the liquid and immersing only the end of the tip, the system benefits from the ultrahigh quality factors ($\sim 10^3-10^4$) associated with the normal and lateral resonance modes at $f_\text{N} = \omega_\text{N}/2\pi\sim 32$ kHz and $f_\text{T}=\omega_\text{T}/2\pi\sim 18$ kHz, respectively. Combined with the large equivalent stiffnesses $K_\text{N}\approx 40$ kN$\cdot$m$^{-1}$ and $K_\text{T}\approx 12$ kN$\cdot$m$^{-1}$, this allows for excellent force sensitivity and ultra-stable, precise control over the tip displacement.

In this configuration, the interaction forces associated with the dynamic excitation of the confined liquid are characterized through the normal and tangential complex mechanical impedances, $Z_\text{N}^*$ and $Z_\text{T}^*$ [N$\cdot$m$^{-1}$], defined as the ratio between the forces and the oscillation amplitude in the normal (N) and tangential (T) directions, respectively. The real part $Z^{\prime}_i\text{ with } i\in\{\text{N},\text{T}\}$ is influenced by conservative interactions, while the imaginary part $Z^{\prime\prime}_i$ accounts for dissipative processes. 

In particular, we refer to the normal conservative impedance $Z^{\prime}_\text{N}$ as the load stiffness, which in the linear regime of small oscillations is physically interpreted as a force gradient, defined as $Z^{\prime}_\text{N} (h)=-\partial F_\text{N} /\partial h$ where $F_\text{N}$ [N] denotes the conservative normal force and $h$ [m] the vertical position of the probe relative to the substrate.

The near-surface forces at play upon water confinement are first reported in Fig.~\ref{fig1:experimentalsetup} c, where we plot the typical evolution of the normal force gradient $Z^{\prime}_\text{N}$, here measured for a tungsten probe approaching the HOPG substrate (see SI Section C for additional data on the other substrates). The data is shown here as a function of the substrate displacement, which due to the high stiffness of the tuning fork is equivalent to the relative position between the probe and the surface. As detailed below, $h_0$ characterizes the onset of hard contact repulsion, associated with 2D water confinement.

Three subsequent regimes are evidenced upon approach, all described in the Derjaguin approximation. As shown in Fig. \ref{fig1:experimentalsetup} d, at large distances from the contact ($D\approx 3-6$ nm), we resolve a negative stiffness that is consistent with attractive van der Waals interactions and yields a Hamaker constant $H = (5 \pm 1)\times10^{-20}$ J, falling within the  range expected from such systems \cite{israelachvili2011intermolecular}. 

Upon further confinement, here $D \in [0,3]$ nm,
an intermediate regime emerges, with the onset of repulsive interactions, characterized by an exponential-like increase of the stiffness with a decay length $\lambda = 0.4\pm 0.1$ nm, as shown in Fig. \ref{fig1:experimentalsetup} e. The molecular scale of this decay length is in line with the onset of repulsive hydration forces \cite{israelachvili2011intermolecular}. Yet, consistent with the absence of resolved oscillatory structural forces, this repulsive profile is also possibly convolved with subnanometric roughness of the probe (more details in SI Section C).

\begin{figure}[htb!]
    \centering
    \includegraphics[width=\textwidth, trim = 1cm 0cm 0cm 1cm, clip]{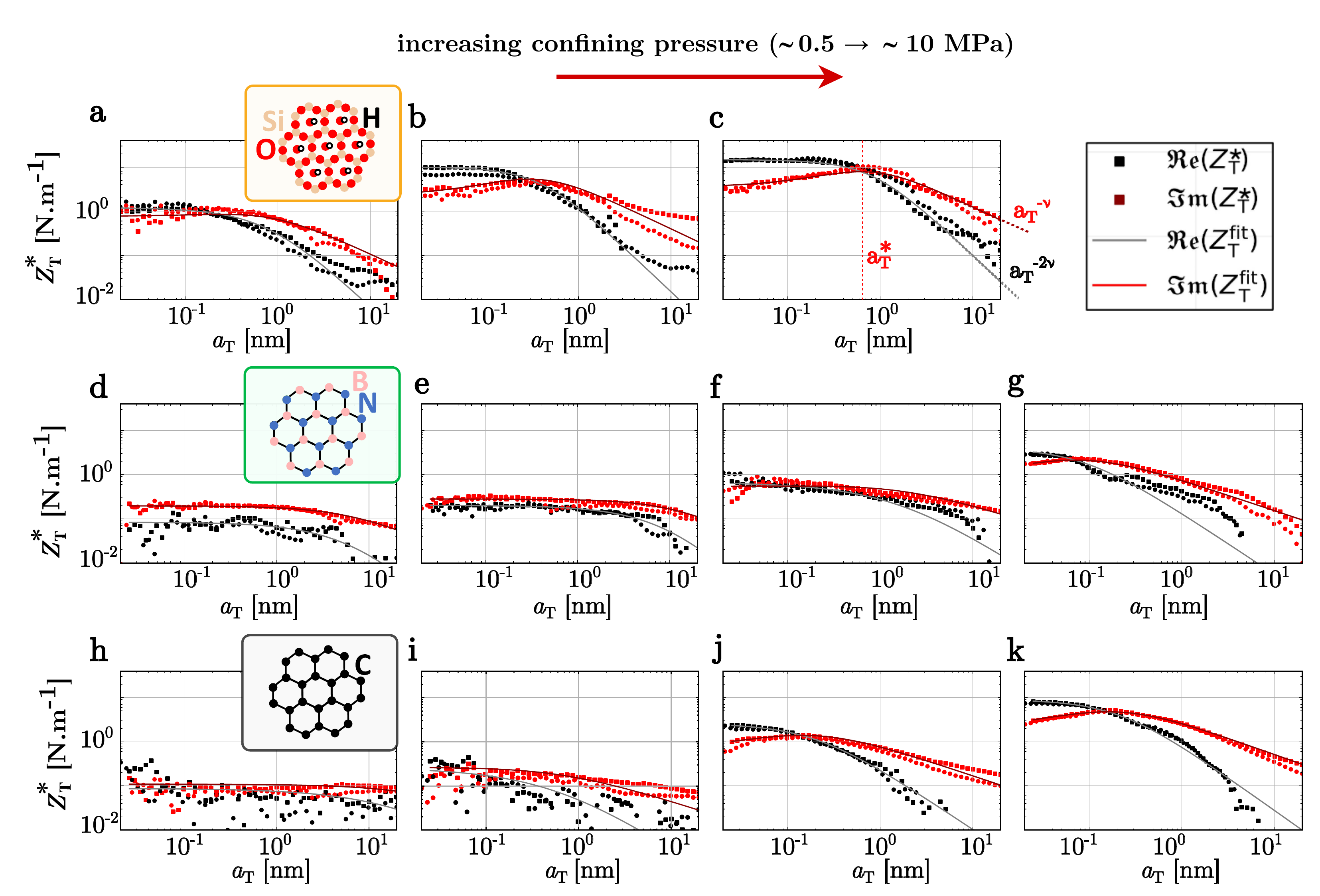}
    \caption{\textbf{Amplitude sweeps in pure water} at up to four load stiffnesses $Z^{\prime}_\text{N}$: 5,  25, 50 and 100 N$\cdot$m$^{-1}$, along with approximate corresponding pressure $P\sim 0.5, 2.5, 5$ and $10$ MPa for \textbf{(a–c)} {Mica}; \textbf{(d–g)} {hBN}; \textbf{(h–k) }{HOPG}. Dark red points: damping $Z^{\prime\prime}_\text{T}$; black points: stiffness $Z^{\prime}_\text{T}$. Squares: increasing amplitude $a_\text{T}$; circles: decreasing amplitude. Plain lines characterize the fit of the lateral mechanical impedance $Z_\text{T}^*$ using the nonlinear viscoelastic model described in Eqs.~\ref{eq:viscoelasticity} and \ref{eq:phenomeno2}, spanning three decades in lateral amplitude $a_\text{T}$ and shown respectively for $Z^{\prime\prime}_\text{T}$ (red) and $Z^{\prime}_\text{T}$ (gray). Characteristic features are highlighted in panel (c), showing the critical amplitude $a_\text{T}^*$ at the crossover $Z^{\prime}_\text{T}=Z^{\prime\prime}_\text{T}$ along with local maximum of $Z^{\prime\prime}_\text{T}$. Below $a_\text{T}^*$ is a solid-like regime and above yielding ($a_\text{T} > a_\text{T}^*$) lies a liquid-like regime, whose asymptotic power laws are also shown as dotted lines.}
    \label{fig2:Amplitude-sweeps}
\end{figure}

Upon further increase of the indentation, we then observe a linear growth of $Z^{\prime}_\text{N}$ with tip displacement (straight solid line, Fig. \ref{fig1:experimentalsetup} c). Such linear growth is consistent with a soft-shell indentation model, here presumably associated with the elastic deformation of a thin layer of WO$_3$ oxide present at the tungsten surface \cite{biegelsen1987ion}. This linear extrapolation allows for a definition of the onset of solid contact $h_0$, and the indentation depth $\delta=h_0-h$. This regime of $h<h_0$ accordingly corresponds to one of solid contact and 2D water confinement. In this regime, control of the indentation $\delta$ leads to the growth of the contact area $A= \pi R\delta$ and the contact pressure $P \sim Z^{\prime}_\text{N}/(2\pi R)$, reaching here up to tens of MPa (Fig. \ref{fig1:experimentalsetup} c, right vertical label). Notably, the mechanical zero position determined from both the van der Waals attraction method and the indentation model agree to within 0.3~nm, demonstrating the consistency of our approach. The crossover from exponential repulsion to linear force growth at $h = h_0$ yields a normal stiffness of $Z^{\prime}_\text{N}\simeq 25$~N$\cdot$m$^{-1}$, corresponding to a pressure of approximately 2.5~MPa above which we assume that the confinement distance saturates at a molecular confinement, estimated as $e_0\simeq 0.5$~nm. Note that comparable force profiles were obtained on the two other substrates, mica and hBN (presented in SI Section C).
\\ \\
\noindent \textbf{Non-linear viscoelasticity.} Following this characterization of near-surface forces, we turn in Fig.~\ref{fig2:Amplitude-sweeps} to the core of our experimental protocol, where we probe the shear-dependent rheological response of water in the confined regimes presented in Fig.~\ref{fig1:experimentalsetup}. Our approach is conceptually analogous to strain-dependent rheological measurements commonly performed on macroscopic samples  \cite{miyazaki2006nonlinear}, whereby the material is submitted to oscillatory strains $\gamma$ of increasing amplitude, while recording its non-linear mechanical response $G^*(\gamma) = G^{\prime}(\gamma)+j\cdot G^{\prime\prime}(\gamma)$. To extrapolate such approaches to ultra-confined water, we fix the normal contact stiffness $Z^{\prime}_\text{N}$, to impose a constant normal pressure and indentation (Fig.~\ref{fig1:experimentalsetup} c). Under these set conditions of geometry and pressure, we progressively increase the lateral tuning-fork oscillatory amplitude, submitting the interstitial liquid to oscillatory shear strains of increasing magnitude.

Figure~\ref{fig2:Amplitude-sweeps} presents these representative amplitude-sweeps on our three lamellar substrates at various imposed normal stiffnesses $Z^{\prime}_\text{N}$, which, using $P\sim Z^{\prime}_\text{N}/2\pi R$, is associated with normal pressures ranging from $\sim 0.5$ to $\sim 10$ MPa. Conceptually analogous to the measurement of shear stress performed in classical rheology, this experimental configuration provides here access to the shear mechanical impedance $Z_\text{T}^* = Z_\text{T}^{\prime} + jZ_\text{T}^{\prime\prime}$ for oscillation amplitude tuned over three decades from 20 pm to 20 nm. The real part $Z_\text{T}^{\prime}$ (black) quantifies interfacial elastic in-phase response, whereas the imaginary part $Z_\text{T}^{\prime\prime}$ (red) accounts for out-of-phase shear dissipation, encompassing both viscous and plastic contributions. 

As a first illustrative case, we focus on Fig.~\ref{fig2:Amplitude-sweeps}~c, associated with dynamic rheology of 2D water on mica, at normal stiffness  $Z^{\prime}_\text{N} = 50$~N$\cdot$m$^{-1}$ corresponding to a pressure $P \sim 5$~MPa. Remarkably, the non-linear response reported here presents features strikingly similar to that associated with the yielding of macroscopic disordered yield-stress fluids and soft glassy materials \cite{bonn2017yield}. We recover in particular:\textit{(i)} a solid-like elastic regime at low oscillation amplitude, where we probe the linear response, with the mechanical impedance dominated by its elastic component ($Z_\text{T}^{\prime} > Z_\text{T}^{\prime\prime}$). \textit{(ii)} a local maximum in the dissipative response $Z_\text{T}^{\prime\prime}$ at intermediate oscillation amplitude, coinciding with the crossing between the elastic and dissipative components ($Z_\text{T}^{\prime} \approx Z_\text{T}^{\prime\prime}$). \textit{(iii)} a fluid-like regime at large oscillation amplitude with a response dominated by dissipation ($Z_\text{T}^{\prime\prime} > Z_\text{T}^{\prime}$), and power-law decays for both $Z_\text{T}^{\prime} \sim a_\text{T}^{-\nu^{\prime}}$ and $Z_\text{T}^{\prime\prime}\sim a_\text{T}^{-\nu^{\prime\prime}}$, where $\nu^{\prime}\approx 2\nu^{\prime\prime}$.

Quantitatively, this non-linear amplitude-dependent response can be fully accounted for within a nonlinear viscoelastic framework  \cite{miyazaki2006nonlinear} characterizing the shear-dependent viscoelastic response of soft yielding materials, phenomenologically combining Maxwell-like response with shear-induced acceleration of the dynamics. In brief, this Maxwell-like response of a fluid with relaxation time $\tau$ [s] (potentially accelerated by shear) and probed at the tuning-fork oscillation frequency $\omega_\text{T}$ [rad$\cdot$s$^{-1}$]  allows us to express the mechanical impedance as
\begin{equation}
Z_\text{T}^* = Z^0_\mathrm{T} \frac{(\omega_\text{T} \tau)^2 + j(\omega_\text{T} \tau)}{1 + (\omega_\text{T} \tau)^2}
\label{eq:viscoelasticity}
\end{equation}
where $Z^0_\mathrm{T}$ [N$\cdot$m$^{-1}$] is the effective lateral stiffness, expressed from the storage modulus as $Z^0_\mathrm{T} = G_0 A/e_0$, with $G_0$~[Pa] the shear modulus, $A$ the contact area and $e_0$ the confined water thickness.

Consequently, the acceleration of the relaxation time of the confined fluid under shear can be accounted for by a phenomenological law for shear-induced fluidification, as:
\begin{align}
    \frac{1}{\tau} = \frac{1}{\tau_0} + k\cdot\dot \gamma^\nu
\label{eq:phenomeno}
\end{align}
where $\tau$ and $\tau_0$~[s] are the relaxation timescales evaluated respectively in the presence and in the absence of shear, and $k\sim O(1)$~[s$^{\nu-1}$] and $\nu\lesssim 1$~[-] phenomenological constants that characterize the nonlinearity of the fluid response.

With the oscillation frequency $\omega_\text{T}$ kept fixed, the control variable in our experimental configuration is the oscillation amplitude $a_\text{T}$, which sets equivalently the shear strain $\gamma = a_\text{T}/e_0$ and shear rate $\dot \gamma = \omega_\text{T}\gamma$ applied to the confined contact,
where $e_0$ is a confinement distance of molecular dimensions. Assuming a confinement thickness of $e_0 = 0.5$~nm associated with two water monolayers, the corresponding shear strains explored in this work span the linear to non-linear range with $\gamma \in [0.04,\,40]$, while the associated shear rates extend from $\dot{\gamma} \in [4\times10^3,\,4\times10^6]$~s$^{-1}$ \textit{i.e.}. We thus express the evolution of the normalized shear-induced relaxation time of our system $\omega_\text{T} \cdot \tau$ as
\begin{equation}
    \frac{1}{\omega_\text{T} \cdot \tau} = \frac{1}{\omega_\text{T} \cdot \tau_0} + \left(\frac{a_\text{T}}{a_\text{T}^*} \right)^\nu
\label{eq:phenomeno2}
\end{equation}
where $a_\text{T}^*$~[m] is the critical shear amplitude above which the dynamics get fluidized, and the product $\text{De} =\omega_\text{T}\tau_0$ can be seen as a Deborah number, characterizing the relative magnitude of the intrinsic (zero shear) relaxation time of the system $\tau_0$ divided by the characteristic time of the excitation probe $1/\omega_\text{T}$.

As shown by the gray and red line fits in Fig. \ref{fig2:Amplitude-sweeps}, this approach allows for a quantitative description of the entire non-linear viscoelastic behavior evidenced experimentally over the full range of contact pressure and substrates -- a truly remarkable result, given the extreme simplicity of the model. Coming back to the case of confined water on mica, Fig. \ref{fig2:Amplitude-sweeps}~c, the yielding-like response evidenced here is characteristic of large Deborah number $\text{De}>1$, associated with long intrinsic (zero-shear) relaxation time $\tau_0>1/\omega_\text{T}\simeq 10$~µs. At low oscillation amplitudes, $Z^{\prime}_\text{T}$ and $Z^{\prime\prime}_\text{T}$ converge towards asymptotic values $Z^{\prime}_\text{T}(a_\text{T}\rightarrow 0)\simeq Z^0_\mathrm{T}$ and $Z^{\prime\prime}_\text{T}(a_\text{T}\rightarrow 0)\simeq Z^0_\mathrm{T}/\text{De}<Z^{\prime}_\text{T}$. The critical amplitude $a_\text{T}^*$ can be identified as point of intersection between $Z^{\prime}_\text{T}$ and $Z^{\prime\prime}_\text{T}$, coinciding also with the maximum in dissipation. At large oscillation amplitude, the liquid-like response of the system is characterized by asymptotic scalings $Z^{\prime}_\text{T}\sim (a_\text{T})^{-2 \nu}$ and $Z^{\prime\prime}_\text{T}\sim (a_\text{T})^{-\nu}$ with here  $\nu \sim 1$. In particular, the non-linear response evidenced here is consistent with fluid relaxation reaching $\tau_0 = 40\pm 2$~µs, in line with $\text{De}\sim 4>1$. At lower load stiffness $Z^{\prime}_\text{N}=5$ N$\cdot$m$^{-1}$ presented in Fig.~\ref{fig2:Amplitude-sweeps}~a, where the confinement distance is estimated at about 1~nm (\textit{cf.} Fig.~\ref{fig1:experimentalsetup}~c), the response is qualitatively the same, but with $\text{De}\gtrsim1$ so just high enough to exhibit yielding.

Keeping the load stiffness at a set value of 5~N$\cdot$m$^{-1}$, we turn towards the non-linear shear-induced response on the second type of substrate, here the hBN crystal on Fig.~\ref{fig2:Amplitude-sweeps}~d. The shape of the dynamic response is qualitatively distinct, with a liquid-like response in the linear low shear amplitude regime (associated to $Z^{\prime}_\text{T}<Z^{\prime\prime}_\text{T}$), transitioning at large amplitudes towards a shear-thinning regime. Yet, the overall response can be again described using the exact same phenomenological approach, associated this time with a Deborah number $\text{De} \sim 0.4  <1$ and $\tau_0 = 4\pm1 $~µs, suggesting a much reduced relaxation time on the hBN substrate compared to mica. In this second low $\text{De}$ regime, the transition to shear thinning kicks in for $a_\text{T}\sim a_\text{T}^*/\text{De}$, accordingly shifting to increasingly larger amplitudes as the relaxation time decreases.

This transition from shear-thinning to yielding can be quantitatively captured by the sole value of the Deborah number, with a transition at $\text{De}\approx 1$. As $Z^{\prime}_\text{T}/Z^{\prime\prime}_\text{T}(a_\text{T}\rightarrow0) \sim \text{De}$, the transition between these two regimes is also associated to a switch from a viscous to an elastically dominated regime in the low amplitude linear response region. The local maximum in dissipation at intermediate amplitude arises solely for $\text{De} >1$, while shear thinning regimes are systematically recovered at high amplitude (see Fig. SI 1).

We can now come back to the response at various loads on the three substrates of mica, hBN and HOPG. On mica, Figures~\ref{fig2:Amplitude-sweeps} a–c show a local maximum in dissipation $Z^{\prime\prime}_\text{T}$ at all loads, indicating that the interfacial water behaves as a yield-stress fluid in the entire explored range, with $\text{De} > 1$. As $Z^{\prime}_\text{N}$ and normal pressure increase, the ratio $Z^{\prime}_\text{T}/Z^{\prime\prime}_\text{T}$ at low amplitudes rises, suggesting slower relaxation. Accompanied by a growth of the elastic plateau, both $Z^0_\mathrm{T}$ and $a_\text{T}^*$ tend to increase with load.

In contrast to mica, the viscoelastic response on hBN and HOPG at low load is predominantly shear-thinning, with dissipation exceeding the elastic response as $Z^{\prime\prime}_\text{T} > Z^{\prime}_\text{T}$. For lower loads ($Z^{\prime}_\text{N} < 50$ N$\cdot$m$^{-1}$), no clear dissipation peak is seen, indicating $\text{De} < 1$. As $Z^{\prime}_\text{N}$ increases, $Z^{\prime}_\text{T}$ and $Z^{\prime\prime}_\text{T}$ start to cross, with  $Z^{\prime}_\text{T} \sim Z^{\prime\prime}_\text{T}$ in the limit of low amplitudes at $Z^{\prime}_\text{N} \simeq 50$ N$\cdot$m$^{-1}$,  consistent with $\text{De} \sim 1$. At higher loads ($Z^{\prime}_\text{N} > 50$ N$\cdot$m$^{-1}$), the system exhibits yield-stress behavior, qualitatively similar to mica, indicating that the relaxation time $\tau_0$ has increased to give $\text{De}>1$.  Yet, we note that the critical amplitude $a_\text{T}^*$ seems to take much lower values than on mica, a point to which we will come back later. Finally, the complex-fluid exponent $\nu$, extracted from fits to Eqs.~\ref{eq:viscoelasticity} and \ref{eq:phenomeno2}, was generally close to unity ($\nu \simeq 1$), with rarer cases exhibiting lower values down to $\nu \simeq 0.7$. Interestingly, this exponent is close to the shear-thinning power-law exponent predicted in sheared supercooled liquids, which is typically around 0.8 \cite{furukawa2009anisotropic}.
\begin{figure}[H]
    \centering
\includegraphics[width=\textwidth, trim = 0cm 4cm 0cm 0cm, clip]{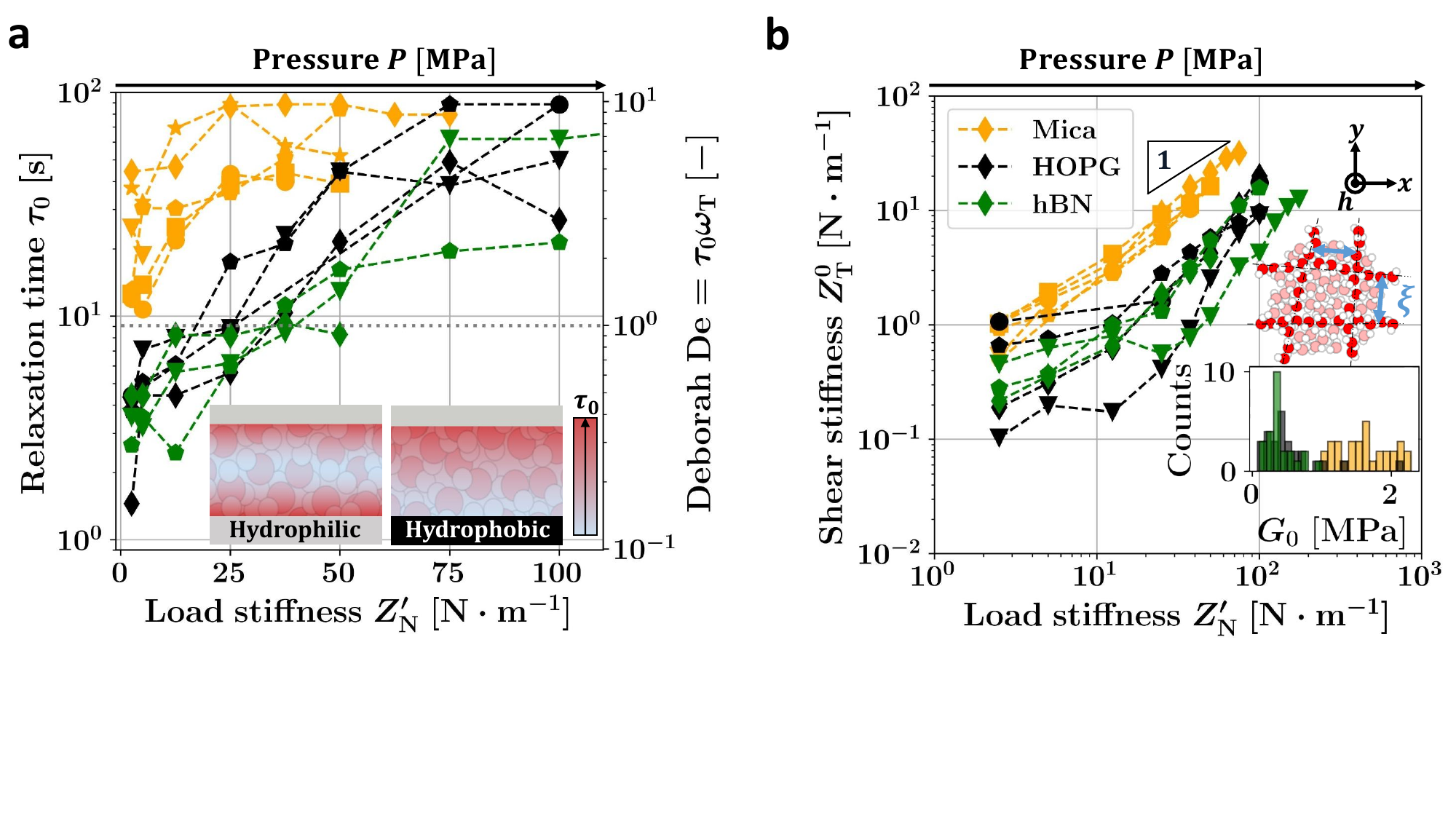}
    \caption{\textbf{Dynamical slowdown of confined water under pressure.}
    \textbf{(a)} {Intrinsic relaxation timescales $\tau_0$ as a function of load stiffness $Z^{\prime}_\text{N}$.} $\tau_0$ is extracted from fits to Eqs.~\ref{eq:viscoelasticity} and \ref{eq:phenomeno2} for tungsten on mica (orange), HOPG (black), and hBN (green) in confined pure water. The dashed horizontal line marks the transition at $\text{De} = \omega_\text{T} \tau_0 = 1$, separating shear-thinning (De $<1$) and yield-stress (De $>1$) regimes. Inset: {Gradient of relaxation times} between (left) two hydrophilic surfaces and (right) a hydrophilic and a hydrophobic surface (Janus confinement) from slow (red) to fast (blue) dynamics.
    \textbf{(b)} {Viscoelastic stiffness $Z^0_\mathrm{T}$.} Fitting parameter describing the magnitude of the shear response. Inset: Histogram showing estimation of the viscoelastic modulus $G_0$ [MPa] and schematics of collective motion in nanoconfined water over nanometric correlation length $\xi$.}
    \label{fig3:relaxationtimescales}
\end{figure}

\noindent \textbf{Dynamical slowdown of confined water.} This non-linear rheological framework allows us to perform a quantitative evaluation of the intrinsic relaxation timescale of the confined fluid, which we report in Fig.~\ref{fig3:relaxationtimescales} a for the three substrates and over the full range of confining pressures. The various symbols of the same color are here associated with distinct experiments, corresponding to a distinct tip/substrate couple. As apparent in this graph, the relaxation time of the interstitial fluid shows a marked increase with pressure, along with a strong substrate dependence. For water confined between the tungsten probe and mica (orange points) the Deborah number $\text{De} = \omega_\text{T} \tau_0$ systematically exceeds unity, consistent with the yield-stress behavior observed at all loads in Fig.~\ref{fig2:Amplitude-sweeps} a--c. While a saturation of $\text{De}$ is observed at high loads, we attribute this effect to the difficulty of evaluating the linear response region at low amplitude when $\text{De} \gg1$, rather than to a physical effect {(see SI Section A).} 
On HOPG and hBN (respectively black and green points), the relaxation time shows a similar rise with increasing pressure, albeit with an order-of-magnitude difference compared to mica. The relaxation progressively transitions from $\text{De} \ll 1$ (shear-thinning, Fig. \ref{fig2:Amplitude-sweeps} d, h) to $\text{De} \gg1$ (yielding, Fig. \ref{fig2:Amplitude-sweeps} g, k). Empirically, the relaxation time is found to increase exponentially with pressure, as $\tau_0\sim \exp(P/P_0)$ with $P_0\approx 2$ MPa, increasing from 1 to 100 µs at the highest pressure $P \sim 10$ MPa. This steady and continuous increase in relaxation time with applied pressure, combined with the characteristic rheological features of soft glassy materials shown in Fig. \ref{fig2:Amplitude-sweeps} is remarkable, suggesting that confined water behaves as a supercooled liquid. The increase in confining pressure leads to a slowdown of the structural relaxation and a progressive shift towards the glass transition, a point we will discuss in further details below. 
\\ \\
\noindent \textbf{Linear and non-linear elastic properties.} Beyond relaxation, we then focus on the evolution of the stiffness $Z^0_\mathrm{T}$ [N$\cdot$m$^{-1}$] characterizing the viscoelastic modulus. This quantity is directly extracted from the rheological data of Fig. \ref{fig2:Amplitude-sweeps}, and reported in Fig.~\ref{fig3:relaxationtimescales} b as a function of the contact stiffness $Z^{\prime}_\text{N}$.  Marked differences are again evidenced between the three substrates with significantly smaller values on the two hydrophobic 2D materials. Expressing this lateral stiffness as $Z^0_\mathrm{T} = G_0 A/e_0$, with $A$ the contact area and $e_0$ the confinement distance, we expect a linear increase of $Z^0_\mathrm{T}$ with the contact area $A=\pi R\delta$ or indentation distance $\delta$ in the high pressure 2D monolayer regime (typically for $Z^{\prime}_\text{N}\gtrsim25$ N$\cdot$m$^{-1}$ or $P\gtrsim 2.5$ MPa) where the confinement distance becomes constant. As $\delta \propto Z^{\prime}_\text{N}$ (Fig. \ref{fig1:experimentalsetup} c), the linear increase of $Z^0_\mathrm{T}$ with $Z^{\prime}_\text{N}$ observed for $Z^{\prime}_\text{N} \gtrsim 25$ N$\cdot$m$^{-1}$ is then consistent with a constant (yet substrate-dependent) value of the shear modulus.  
Taking $e_0 \approx 0.5$ nm, corresponding to a two-monolayer-thick water film, we obtain a shear modulus respectively of $G_0 = 1.7 \pm 0.4$ MPa, $0.5 \pm 0.3$ MPa and $0.4 \pm 0.2$ MPa on mica, HOPG and hBN (see Fig. \ref{fig3:relaxationtimescales} b Inset). 
These values lie in the "soft" MPa range -- three orders of magnitude below the bulk high frequency modulus of water \cite{eichler2025shear} -- tentatively expressed as $G_0\sim k_B T / \xi^3$, where $\xi$ denotes the characteristic correlation length that sets the size of the energy-storing domain. In classical soft disordered materials, this value corresponds for example to the mesh size in a polymeric network or particle size for a colloidal glass \cite{de1979scaling}. 
In our system, $\xi$ is found to be of the order of a few nanometers, suggesting the presence of elastic correlations -- possibly related to dynamical heterogeneities -- extending over domains much larger than the molecular size, and found larger on the two hydrophobic 2D materials.

We can then focus on the yielding behavior of the confined glassy interface, \textit{i.e.} its transition from elastic deformation to flow. Yielding can here be characterized by \textit{(i)} the critical amplitude $a_\text{T}^*$, associated to a yield strain $\gamma^* = a_\text{T}^*/e_0$, and \textit{(ii)} the yield force $F_\mathrm{Y} = Z^{\prime}_0 \times a_\text{T}^*$ or yield stress $\sigma_\text{Y} = F_\mathrm{Y}/A$, where $Z^{\prime}_0$ is the low-amplitude elastic plateau. For clarity, we restrict our analysis in the following to the yielding behavior at high $\text{De}$, for which the critical amplitude $a_\text{T}^*$ is unambiguously identified. 
On mica, this yielding amplitude is systematically found around $a_\text{T}^*\approx 0.3$ nm, and shows a weak increase with pressure. This value is of the same order as the size of a water molecule ($\simeq0.25$ nm), associated with a critical yielding deformation $\gamma^*\approx 0.6 = O(1)$, and a yield stress $\sigma_\text{Y}$ in $1$ to $10$ MPa range (presented in SI Section D).
However, for the two hydrophobic HOPG and hBN materials, fluidization amplitudes are generally lower and more scattered, reaching in some cases values as low as a few tens of picometers (Fig. SI 6), much smaller than typical molecular dimensions. Drawing analogy with the shear banding evidenced in macroscopic disordered glassy systems, we speculate that these low yielding amplitudes are a signature of localization, associated \textit{e.g.} with nucleation of collective slip of the glassy fluid at the solid surface. 
The yield stresses for these conditions are accordingly smaller, reaching $\sigma_\text{Y} = 0.1 - 1$ MPa.
The variability in $a_\text{T}^*$ is then likely related to subtle differences in surface state, such as lattice orientation, uncontrolled in the present experiments. 

We note finally that despite the pronounced non-Newtonian behavior, the energy dissipation can be accounted for in a general sense as a structural viscosity $\eta$~[Pa$\cdot$s], expressed classically for viscoelastic materials as $\eta = \tau_0 G_0$, with $\tau_0$ the relaxation time and $G_0$ the shear modulus. Based on the order of magnitude evaluation described above, interfacial water viscosity would reach values of the order of $\eta \sim 1$--$10$~Pa$\cdot$s for hBN and HOPG and $\sim 10-100$~Pa$\cdot$s for mica, corresponding to an enhancement of $10^3-10^{5}$ relative to bulk water.
\\ \\
\noindent \textbf{Discussion.} The rich phenomenology evidenced here for nanoscale rheological properties for confined and pressurized water exhibits striking features reminiscent of soft glassy rheology, typically associated with disordered materials such as foams, emulsions, and granular systems. Key features associated with these soft glassy materials are their structural disorder and metastable character, leading to flow behaviors reminiscent of glassy or supercooled liquids -- properties which are thus expected to extend to the confined and pressurized water phase under investigation in this work. The continuous increase in relaxation time with pressure is accordingly associated with a continuous slowdown of water dynamics upon increasing pressure, due to reduced molecular mobility. 
In particular, these observations are characteristic of heterogeneous and disordered phases, at odds with a true freezing transition, which has been put forward in some work for water under confinement \cite{giovambattista2009phase,chen2016two, algara2015square} or under bulk pressurization \cite{lee2026multiple, kapil2022first}.

In this respect, it is interesting to put our rheological analysis in perspective with  nanotribological models \cite{persson2013physics}, traditionally used in the context of sliding friction, or plasticity. 
These approaches are built upon shear-facilitated, thermally activated hops based on Eyring models. In these scenarios, dissipation occurs upon overcoming the energy barriers of the corrugated surface potential, with energy loss stemming from phonons degrading into heat during the barrier relaxation process. In particular, periodic corrugated potentials, associated with well-defined energy barriers and a corrugation length, are classically considered, leading to a logarithmic scaling of the friction force with speed, which has been reported in previous experimental boundary lubrication settings, ranging from ion-induced hydration in SFA \cite{ma2015origins}, to water mediated graphene/water lubrication \cite{diao2019slippery}.

Interestingly, approximately logarithmic trends can also be evidenced in our data, when plotting the dissipative force $F_\text{T}^\text{D} =Z^{\prime\prime}_\text{T}\cdot a_\text{T}$ as a function of the harmonic drive velocity $v_\text{T}\sim a_\text{T}\omega_\text{T}$ (detailed in SI Section E). Yet this representation also highlights a critical missing feature of these models, in the form of an over-dissipation at intermediate velocities, leading to deviations from purely logarithmic scaling and underscoring the limitations of these simple barrier-hopping models.
In physical terms, this local maximum in dissipative impedance is a hallmark of the collective structural relaxation at play in glassy systems and stems from the decrease of the structural relaxation time with the applied shear rate, bringing it closer to the characteristic excitation frequency, a point at which dissipation becomes maximal \cite{wyss2007strain}.
Note that more advanced frameworks, based, \textit{e.g.}, on Mode Coupling Theory \cite{miyazaki2006nonlinear}, can be used to reproduce formally these phenomenological results with key ingredients stemming from metastability and disorder.

Coming back to boundary lubrication, previous results \cite{ma2015origins,diao2019slippery} have been interpreted as stemming from a confinement-induced increase in the corrugation potential, leading to a transition from viscous-like thermally dominated transport to a logarithmic-like, force-induced barrier hopping. Yet these transitions also resemble qualitatively the smooth pressure-induced transition from low to high Deborah number evidenced here, which is rather associated to a dynamic slowdown. Reinterpreting these data within this non-linear rheological framework and reconciling these two visions presents exciting perspectives. 
All in all, the universality of these rheological responses, extending from macroscopic pasty rheology to nanoconfined water is truly remarkable. Exploring the specificity associated to the 2D nature of the system under investigation would be of high interest.  

We finally come back to the marked substrate difference evidenced in our experiments, pointing towards distinct interfacial relaxation modes for water. The slowdown is most pronounced for the hydrophilic mica substrate, which can form hydrogen bonds with water \cite{trushin2025structure}. However, we find faster yet comparable relaxation times for the HOPG and hBN substrates. This indistinguishability is surprising: despite sharing similar crystallography, several reports have indeed demonstrated strong differences in water friction and interaction \cite{wu2024probing,secchi2016massive,keerthi2021water}, at the root of which was argued to be their electronic character, with graphite a semi-metal and hBN a wide band gap insulator. This points towards hydrophobicity as the dominant mechanism governing water dynamics in the gap \cite{huang2008water}, suggesting that relaxation time and slip are governed by distinct mechanisms \cite{herrero2020fast, carlson2025subnanometer,schlaich2017hydration}. As presented in the inset of Fig.~\ref{fig3:relaxationtimescales} a, considering that the surface state of the tungsten probe is likely hydrophilic due to tungsten oxides, experiments on mica would be associated with symmetric hydrophilic confinement and relatively homogeneous relaxation time at the two interfaces. In contrast, for the Janus confinement, \textit{i.e.} between hydrophilic and hydrophobic walls, facilitated relaxation dynamics could arise at the hydrophobic interface.

The non-linear response evidenced here thus brings exciting perspectives for our understanding of confined water dynamics. While the present investigation was restricted to a single shear frequency in the kHz range, imposed by the resonance frequency of the QTF tangential mode, the relaxation dynamics of nanoconfined liquids, however, are likely distributed over a broad range of timescales and may involve multiple relaxation processes. Likewise, the non-linear transition evidenced for shear-dependent responses could also have a signature in the frequency domain. Clarifying these effects by measuring frequency- and amplitude-dependent responses \cite{wyss2007strain}, through superposition rheometry \cite{comtet2019atomic} or in-situ dielectric spectroscopy \cite{fumagalli2018anomalously,wang2025plane} could provide a powerful way of probing the full relaxation spectrum and nonlinear response over extended timescales, to clarify the underlying rheological behavior. Along these lines, recent approaches have coupled various spectroscopic modalities, such as Raman \cite{hwang2024nano}, NMR \cite{pagliero2025slow} or single molecule microscopy \cite{ronceray2023liquid} based on color centers or non-linear optical generation \cite{dhopatkar2016ice} to reveal structural transitions in confined and interfacial water, some of them coupled with scanning probe microscopy \cite{hwang2024nano}. Such couplings could be fruitful by providing direct access to possible shear-induced dynamical heterogeneity in nanoconfined water and could reveal how structural rearrangements correlate with the observed rheological response, possibly extending to much higher pressures where highly exotic water phases have been predicted to arise \cite{kapil2022first}. Finally, in the context of fluctuations or quantum-induced friction \cite{kavokine2022fluctuation}, the evidence of this relaxation time shift under confining pressure brings interesting perspectives for engineering and controlling fluctuation-induced dissipation \cite{lizee2024anomalous}, by playing on both liquid and internal wall dynamics.

\section*{Experimental methods}

\noindent\textbf{Instrumentation.} Dynamic atomic force microscopy (AFM) measurements were performed using a quartz tuning fork (QTF) as an electromechanical resonator, excited by a piezodither at resonance frequencies $f_\text{N} \sim 32$ kHz (normal mode) and $f_\text{T} \sim 18$ kHz (tangential mode). The piezodither driving voltage is referred to hereafter as the excitation. Oscillation amplitudes $a_\text{N}$ and $a_\text{T}$ were held constant through independent PID and PLL servo loops (Nanonis Mimea). When conservative forces act on the tuning fork, its resonance frequency shifts by $\delta f_i$ [Hz] relative to a reference state ($i \in \{\text{N, T}\}$), with $\delta f_i > 0$ for repulsive interactions and $\delta f_i < 0$ for attractive ones; when dissipative forces are present, the excitation $E_i$ [V] required to maintain constant oscillation amplitude increases relative to its reference value $E_{i0}$ [V]. Combining both channels, the complex mechanical impedance of the confined liquid is obtained for each mode as
\begin{equation}
    Z_i^* = Z^{\prime}_i+jZ^{\prime\prime}_i= 2K_i\frac{\delta f_i}{f_i}
    + j\,\frac{K_i}{Q_{i0}}\left(\frac{E_i}{E_{i0}} - 1\right)\label{eq:defZ}
\end{equation}
where $K_i$ is the tuning-fork stiffness in the corresponding direction ($K_\text{N} = 40$ kN$\cdot$m$^{-1}$, $K_\text{T} = 12$ kN$\cdot$m$^{-1}$), $f_i$ the resonance frequency, and $Q_{i0}$ the quality factor in the reference state (further calibration details in SI Section B).

\noindent\textbf{Probe fabrication.} A tungsten wire was attached to one prong of the tuning fork using silver epoxy (EPO-TEK H31D). To shape the wire into a sphere and reduce its length, it was dipped into a 1~M sodium hydroxide solution held in a metal ring; applying a 5~V bias induces electrolytic corrosion of the tungsten along the meniscus, forming a spherical tip. Tip radius of curvature $R$ (typically $R \sim 1$ µm, ranging $R \in [0.4, 6]$ µm across probes) and surface smoothness were characterized by optical and electron microscopy. Fits performed at both extremes of this radius range gave results consistent with the rest of the dataset, supporting the robustness of the model.

\noindent\textbf{Sample preparation.} Mica (Ted Pella, Grade V1) and HOPG (Ted Pella, grade ZYB), both available as centimeter-scale crystals, were freshly exfoliated with adhesive tape and immersed within approximately one minute in a custom PEEK liquid reservoir containing $\sim 3$ mL of ultrapure water (MilliQ, 18 M$\Omega\cdot$cm), minimizing airborne contamination. hBN crystals, smaller in lateral size ($\sim 100$ µm to $\sim 1$ mm), were exfoliated and deposited on UV--O$_3$ plasma-activated silicon wafers prior to immersion. Once deposited on silicon, the hBN flakes' typical width was $\sim 10-100$ µm and their thickness $\sim 0.1-1$ µm.

\noindent\textbf{Experimental environment.} A glass cover placed over the reservoir, with an opening left for the probe, limited water evaporation and pinned the meniscus to reduce drift of the triple line. The probe was coarsely positioned with stepwise 3D motors (Mechonics) and finely adjusted with a 3D piezo-scanner (40-µm range, Piezosystem Jena). To preserve high quality factors ($Q_\text{N}, Q_\text{T}$ of several thousand), only the tip of the probe was immersed in the water meniscus connecting the glass slide to the substrate: the tuning fork itself, and the epoxy joint, remained dry throughout. Mechanical isolation from external vibrations was provided by a heavy marble table combined with an air-suspension table beneath the AFM.

\noindent\textbf{Measurement protocol.} Mechanical contact with the hydrated substrate was identified from force-gradient curves (Fig.~\ref{fig1:experimentalsetup}~c), where the soft-shell and hydration repulsion regimes were characterized after allowing piezo-scanner creep and thermal drift to relax. With the load held at constant stiffness $Z^{\prime}_\text{N}$ [N$\cdot$m$^{-1}$] through a PID servo loop, the confinement distance $D$ and confining pressure $P$ were kept constant throughout each amplitude sweep, including at the largest tangential amplitudes explored ($a_\text{T} \le 20$ nm). Amplitude sweeps were recorded over $a_\text{T} = 20$ pm--20 nm at several fixed loads to probe the different amplitude-dependent regimes described in the main text. The analysis takes into account the possibility of a small elastic offset in the shear response (further protocol details in SI Section B).

\noindent\textbf{Reproducibility.} Two complementary criteria were used to identify clean, uncontaminated interfaces. First, only conditions showing force-gradient curves that superpose well on approach and retract were considered. In addition, contact detection had to be sufficiently sharp, with a slope $|\partial Z^{\prime}_\text{N}/\partial h(h\simeq h_0)|>10 $ (N$\cdot $m$^{-1}$)/nm. Second, amplitude sweeps showing large hysteresis or poor reproducibility between increasing- and decreasing-amplitude branches were discarded, as these indicate contamination or non-steady-state regulation; only sweeps with minimal hysteresis, consistent with a clean surface, were retained for the analysis presented in the main text. Applying these criteria, independent experiments -- each corresponding to a distinct probe/substrate pair -- were compiled for five samples on mica, four on HOPG, and three on hBN, spanning different cleaved substrates, probe orientations, and batches of ultrapure water. Residual variability among these experiments is attributed to differences in substrate orientation, probe roughness, or trace surface contaminants such as alkane molecules \cite{hurst2018adventitious}.

\section*{Acknowledgments}

G.M. acknowledges fruitful discussions with C. Cramail and M. Lizée.

J.C. acknowledges funding from the ANR (grant ‘GUACAmole’ ANR-22-CE06-0003-01). This project has also received financial support from the Ile-de-France Region in the framework of ‘DIM Respore’ and ‘DIM MaTerRE’, the CNRS through the MITI interdisciplinary programs and from the Carnot Institute ’IPGG Microfluidique’. This work benefited from the technical contribution of the joint service unit CNRS UAR 3750 and was supported by UAR 2063 Workshop for Advanced Research Support.  K.W. and T.T. acknowledge support from the CREST (JPMJCR24A5), JST and World Premier International Research Center Initiative (WPI), MEXT, Japan.

\renewcommand\refname{References}
\bibliography{references}

\end{document}


\appendix

\renewcommand{\thefigure}{SI.\arabic{figure}} 
\setcounter{figure}{0} 

\renewcommand{\thetable}{SI.\arabic{table}} 
\setcounter{table}{0} 

\renewcommand{\figurename}{Supplementary Figure}

\renewcommand{\tablename}{Supplementary Table}
\maketitle
email: jean.comtet@espci.fr
\tableofcontents 




\newpage

\section{Modeling of the dynamic viscoelastic response}
\subsection{Main ingredients.}

We recall below the main ingredients of our amplitude-dependent viscoelastic dissipation model, adapted from bulk viscoelastic dissipation laws, developed for example in Ref.~\cite{miyazaki2006nonlinear}.

\textit{(i)} a Maxwell-like response of the material, characterized by a spring and a dashpot connected in series (Eq.~\ref{eq:viscoelasticity}).

\begin{equation}
    Z_\text{T}^* = Z^0_\mathrm{T}\, \frac{(\omega_\text{T} \tau)^2 + j(\omega_\text{T} \tau)}{1+(\omega_\text{T} \tau)^2}\label{eq:viscoelasticity}
\end{equation}
where $Z^0_\mathrm{T}$ [N$\cdot$m$^{-1}$] corresponds to the stiffness of the elastic spring, $\omega_\text{T}=2\pi f_\text{T}$ [rad/s] is the angular frequency of the tangential resonance mode, and $\tau$ [s] is the effective relaxation timescale of the fluid, potentially accelerated by shear. Here, $j$ denotes the imaginary unit ($j^2=-1$). 

\textit{(ii)} In parallel, a phenomenological law is introduced to account for the shear-induced fluidification (\textit{cf.} Ref.~\cite{miyazaki2006nonlinear}):
\begin{align}
    \frac{1}{\tau} = \frac{1}{\tau_0} + k\cdot\dot \gamma^\nu.
\label{eq:phenomeno}
\end{align}

Here, $\tau_0$ [s] is the intrinsic relaxation time in the absence of shear and $\dot{\gamma}$ [s$^{-1}$] refers to the shear rate. The exponent $\nu \approx 1$ characterizes the nonlinearity of the fluid response at large shear amplitude. The constant parameter $k$ [s$^{\nu-1}$] usually scales as $k\sim O(1)$  according to Ref.~\cite{miyazaki2006nonlinear}.

The shear rate can be expressed as $\dot \gamma = \omega_\text{T} \cdot a_\text{T}/e_0$ with $\gamma=a_\text{T}/e_0$ [-] the shear strain, $e_0$ the confinement distance of molecular size and $a_\text{T}$ the experimentally controlled shear amplitude.

We thus rewrite:
\begin{equation}
    \frac{1}{\omega_\text{T} \tau} = \frac{1}{\omega_\text{T} \tau_0} + \left(\frac{\gamma}{\gamma^*} \right)^\nu = \frac{1}{\omega_\text{T} \tau_0} + \left(\frac{a_\text{T}}{a_\text{T}^*} \right)^\nu
\label{eq:phenomeno2}
\end{equation}

where $\gamma^*$ and $a_\text{T}^*$ are respectively a critical shear strain and shear amplitude over which the dynamics becomes fluidized, with
\begin{equation}
a_\text{T}^* = \frac{e_0}{k^{1/\nu}\omega_\text{T}^{(\nu-1)/\nu}} \approx e_0 ~\text{and}~ \gamma^* \approx 1 \text{ when } \nu\approx1
\end{equation}

The transition between the shear-thinning and shear-thickening response is set by the ratio of the intrinsic fluid relative relaxation $\tau_0$ and oscillation time-scale $1/\omega_\text{T}$, which we write as a Deborah number:
\begin{equation}
    \text{De} = \tau_0 \omega_\text{T}
\end{equation}

\subsection{Non-dimensionalization.}

Expressing $\tilde a = a_\text{T}/a_\text{T}^*$ and $\tilde Z_\text{T}=Z_\text{T}^*/Z^0_\mathrm{T}$, we write the effective response as:
\begin{equation}
    \tilde Z_\text{T} = \frac{(\text{De}^{-1}+\tilde a^\nu)^{-2} + j(\text{De}^{-1}+\tilde a^\nu)^{-1}}{1+(\text{De}^{-1}+\tilde a^\nu)^{-2}}\label{eq:Dimensionless-viscoelasticity}
\end{equation}
We can express the asymptotic behavior when $\tilde a \rightarrow 0$, leading to:
\begin{equation}
\tilde Z_\text{T} = \frac{\text{De}^2+j \text{De}}{1+\text{De}^2}
\end{equation}
which would be expected of a linear viscoelastic fluid. In this low amplitude regime, the ratio of elastic to dissipative impedance yields:

\begin{equation}
\frac{\tilde Z_\text{T}^{\prime}}{\tilde Z_\text{T}^{\prime\prime}} = \frac{Z_\text{T}^{\prime}}{Z_\text{T}^{\prime\prime}}= \text{De}
\end{equation}

\begin{figure}[htb!]
    \centering
    \includegraphics[width=17.6 cm, trim = 0cm 4cm 0cm 0cm, clip]{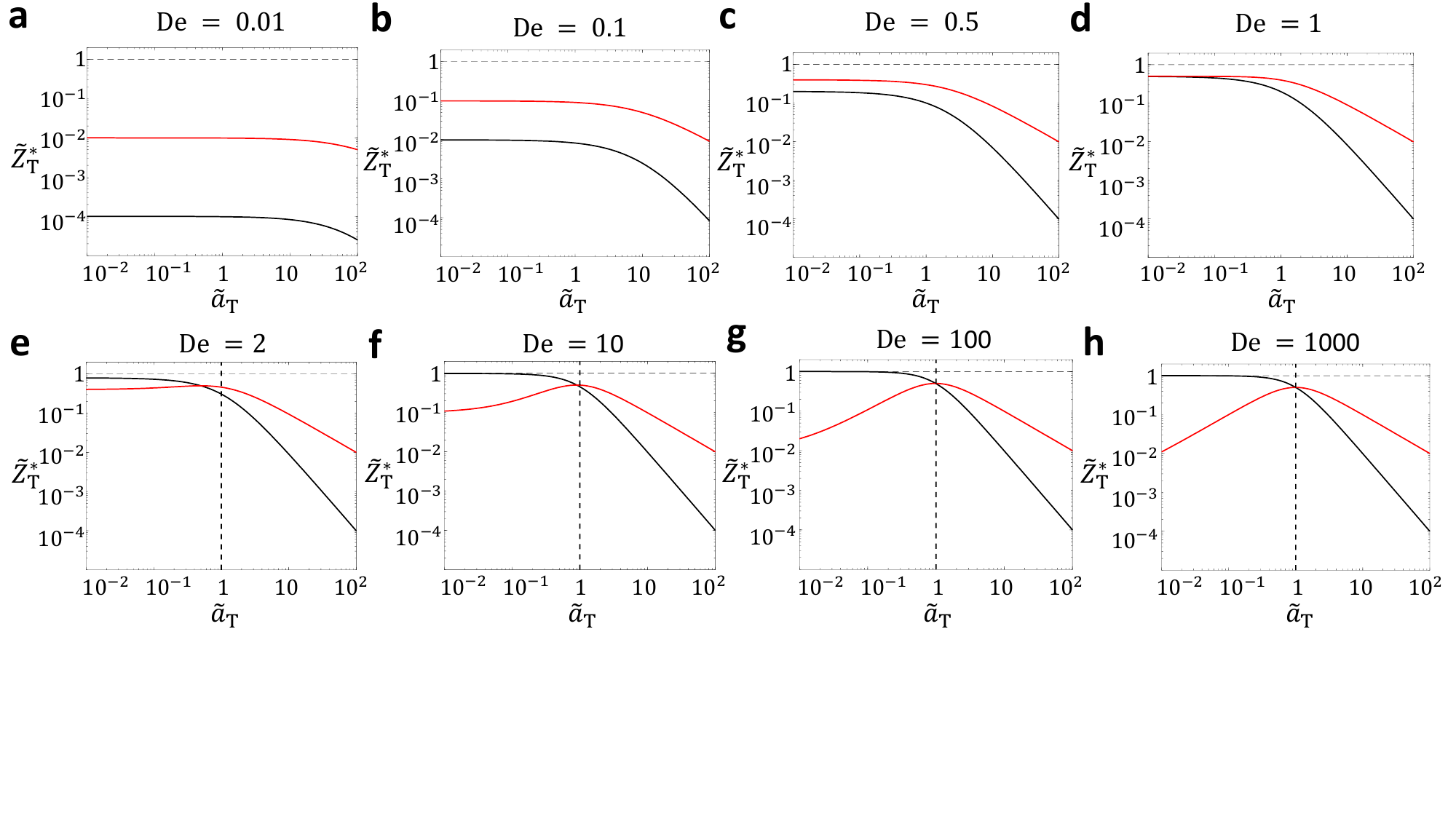}
    \caption{Expected evolution of the mechanical impedance $\tilde{Z}_\text{T}$, as a function of the dimensionless oscillation amplitude $\tilde a=a_\text{T}/a_\text{T}^*$. The real and imaginary parts of $\tilde{Z}_\text{T}$ are shown respectively in black and red. The response is shown for increasing Deborah numbers, respectively in regimes of shear-thinning ($\text{De}\leq 1$, a--d upper row) and yield-stress fluid ($\text{De}>1$, e--h lower row). The horizontal dashed line ($\tilde{Z}_\text{T}= 1$) shows that for $\text{De}>1$, in the low amplitude regime $\tilde{Z}^{\prime}_\text{T}\simeq 1$. The vertical dashed line ($\tilde a=1$) shows that at the yielding amplitude, there is a maximum for  $\tilde{Z}^{\prime\prime}_\text{T}$ and a crossing $\tilde{Z}^{\prime}_\text{T}\simeq\tilde{Z}^{\prime\prime}_\text{T}$.}
    \label{fig:scaling}
\end{figure}

Numerical plots of the response is shown in Fig. \ref{fig:scaling}, where we report $\tilde{Z}_\text{T}=Z_\text{T}^* /Z^0_\mathrm{T}$ as a function of $\tilde a=a_\text{T}/a_\text{T}^*$.

\subsection{Effective response.}

\paragraph{Low Deborah number \textbf{$\text{De}<1$.\\}}
For $\text{De}<1$, we recover a liquid-like shear-thinning response, with $Z_\text{T}^{\prime}<Z_\text{T}^{\prime\prime}$ at low shear amplitude. There is no overshoot in dissipation but a shear thinning behavior at large enough amplitudes. In the limit of $\text{De}\rightarrow 0$, an almost purely Newtonian behavior is recovered, with an amplitude-independent behavior, associated with $Z_\text{T}^{\prime} \sim Z^0_\mathrm{T}\times \text{De}^2 \rightarrow0$ and $Z_\text{T}^{\prime\prime} \sim Z^0_\mathrm{T}\times \text{De}$. 

We also observe on Fig.~\ref{fig:scaling} that as $\text{De}$ decreases, the transition to shear-thinning is shifted towards higher oscillation amplitudes. This trend can be rationalized by noting that the transition from the plateau region to shear-thinning occurs when $\tilde a^\nu \sim \text{De}^{-1}$ \textit{i.e.} at $\tilde a\sim \text{De}^{-1/\nu}$. As $\nu\sim 1$, this is consistent with the shift of the shear-thinning region to the higher amplitudes when $\text{De}$ decreases.

Therefore, if one fits only $Z_\text{T}^{\prime}(a_\text{T})$ or $Z_\text{T}^{\prime\prime}(a_\text{T})$ without the other part, the independent determination of $a_\text{T}^*$ and $\text{De}$ as fitting parameters can become impossible. Moreover, a small systematic error of measurement on $Z_\text{T}^{\prime}$ can lead to significant errors on the independent estimation of $a_\text{T}^*$ and $\text{De}$. Consequently, the experimental setup was assessed to be sensitive to $\text{De}$ in the range $\text{De}\gtrsim0.1$.

\paragraph{High Deborah number \textbf{$\text{De}>1$.\\}}
For $\text{De}>1$, we recover a solid-like yielding response, with the following regions:
\begin{itemize}
    \item \textbf{Plateau region for} $\tilde a_\text{T}\lesssim \text{De}^{-1/\nu}$: we find the constant values $\tilde{Z}^{\prime}_\text{T}\sim \text{De}^2/(1+\text{De}^2)$ and $\tilde{Z}^{\prime\prime}_\text{T}\sim \text{De}/(1+\text{De}^2)$. The response is solid-like as $Z^{\prime}_\text{T}>Z^{\prime\prime}_\text{T}$.
    \item \textbf{Shear-thickening for} $\text{De}^{-1/\nu}\lesssim \tilde a_\text{T}\lesssim 1$: the response is still solid-like but the dissipation increases with $\tilde{a}_\text{T}$.
    \item \textbf{Yielding at} $\tilde{a}_\text{T}\approx 1$: this is an inflection point for stiction and a maximum for dissipation. In addition, the real and imaginary parts are comparable with $\tilde{Z}^{\prime}_\text{T}\sim\tilde{Z}^{\prime\prime}_\text{T}\sim1/2$.
    \item \textbf{Shear-thinning for} $\tilde{a}_\text{T}> 1$: the response is liquid-like as $Z^{\prime}_\text{T}<Z^{\prime\prime}_\text{T}$ and progressively reaches its asymptotic power laws $\tilde{Z}^{\prime}_\text{T}\sim(\tilde{a}_\text{T})^{-2\nu}$ and $\tilde{Z}^{\prime\prime}_\text{T}\sim(\tilde{a}_\text{T})^{-\nu}$.
\end{itemize}

In this yield-stress regime, the estimation of $a_\text{T}^*$ is very precise as there is no influence of $\text{De}$ on the location of the dissipation maximum. However, the precise determination of $\text{De}$ can become challenging when the plateau region is pushed further towards the lower bound of accessible amplitudes. Therefore, the experimental range of accessible amplitudes being $a_\text{T}\in [0.02, ~20]$ nm and as critical amplitudes usually stand in the range $a_\text{T}^*\sim 0.1$--1 nm, we are sensitive to $\text{De}\lesssim 10$. Indeed, for slower relaxation $\text{De}\gtrsim10$, the plateau region is shifted out of the accessible range (below 20 pm).

\paragraph{Experimental limitations}
In conclusion, the experimental setup is able to capture the nonlinear viscoelastic rheology of confined liquids with optimal sensitivity when $\text{De}\in [0.1,~10]$ \textit{i.e.} for $\tau_0\in [1, ~100]$ µs.

\newpage
\section{Data acquisition and analysis}
\subsection{Reference calibrations}

\begin{table}[h!]
    \centering
    \begin{tabular}{|l|c|c|}
        \toprule
    
         Parameter & Normal (tapping) mode & Tangential (shear) Mode \\
        \midrule
         
         Resonance frequency $f_0$ [kHz] & 31--32 & 17--18 \\
         Quality factor in air $Q_0$ [-] & 10,000--13,000 & 1,000--8,000\\
         Effective stiffness $K$ [kN$\cdot$m$^{-1}$] & 40 & 12  \\
         Piezoelectric coupling $\alpha$ [nA$\cdot$nm$^{-1}$] & 2.0 & 0.034\\
         
        \bottomrule
         
    \end{tabular}
    \caption{Electromechanical parameters of the QTF. Calibration values from \cite{rodrigues2009bringing,nigues2012manipulation}.}
    \label{tab:CalibrationQTF}
\end{table}

The proportionality factor $\alpha$---presented in Table~\ref{tab:CalibrationQTF}---was calibrated in a previous work using topography measurements~\cite{rodrigues2009bringing}. For increasing excitation voltage, the QTF current was measured at solid--solid contact together with the piezo-scanner displacement. An increase in oscillation amplitude $\Delta a$ corresponds to a higher apparent contact position $\Delta h$, such that $\Delta a = \Delta h$. A linear relationship between displacement and current was observed, and the calibration factor $\alpha$ was determined from a linear regression. Identical QTFs (reference LF XTAL002995, IQD) were used throughout the present work to ensure consistency in the calibration parameters.

The effective stiffness was also calibrated in a previous work~\cite{nigues2012manipulation}, using an AFM cantilever of known stiffness $K_\mathrm{cantilever}$ exerting a normal force on the QTF. The resulting force gradient shifts the resonance frequency by $\delta f$, such that $K_\mathrm{cantilever} = 2K\,\delta f / f_0$, allowing determination of the effective stiffness $K$.

\subsection{Details on amplitude-sweep acquisition}

\paragraph{Reference calibration.} We first perform the calibration of the excitation as a function of shear amplitude to account for QTF nonlinearity.

\begin{figure}[htb!]
    \centering
    \includegraphics[width= 15 cm, trim = 0cm 8cm 0cm 0cm, clip]{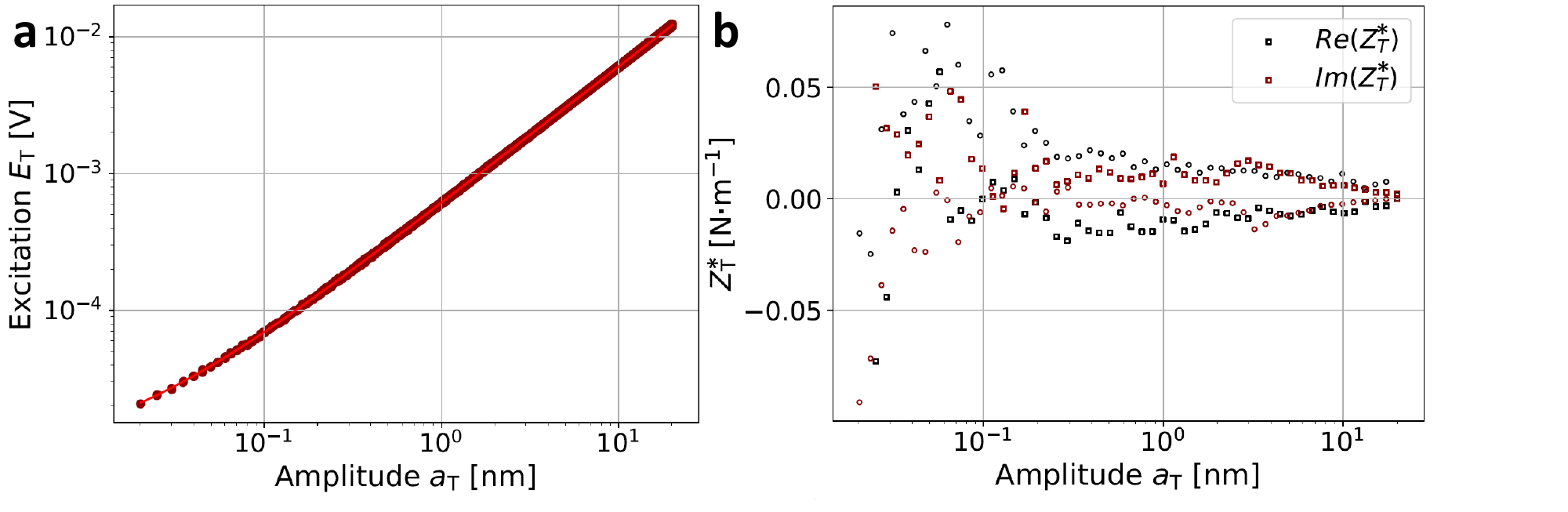}
    \caption{\textbf{Amplitude sweep calibration:} \textbf{(a)} The excitation voltage $E_\text{T}$ [V] (red dots) is fitted by an affine function (red line) of the shear amplitude $a_\text{T}$ [m]. \textbf{(b)} Reference tangential impedance ${Z}^*_\text{T}$ [N$\cdot$m$^{-1}$] measured far from the surface as a function of $a_\text{T}$. Circles: increasing $a_\text{T}$. Squares: decreasing $a_\text{T}$. Black: conservative component $Z^\prime_\text{T}$. Red: dissipative component $Z^{\prime\prime}_\text{T}$.}
    \label{fig:AmplitudeSweepsCalibration}
\end{figure}

Forward and backward $a_\text{T}$-sweeps are first performed far from contact to obtain calibration curves. As shown in Fig.~\ref{fig:AmplitudeSweepsCalibration}(a), the excitation voltage of the tangential mode $E_\text{T}$ [V] is well described by an affine relation:
\begin{equation}
    E_\text{T}(a_\text{T}) = E_{\text{T}0} + \beta_\text{T} a_\text{T}
\end{equation}
where $\beta_\text{T}$ [V$\cdot$m$^{-1}$] typically converts the excitation of the piezo dither into amplitude of lateral oscillation of the QTF. The offset term $E_{\text{T}0}\sim 0.01$~mV accounts for internal damping within the tuning fork.

When properly calibrated, the reference curves shown in Fig.~\ref{fig:AmplitudeSweepsCalibration}(b) are flat and centered around zero. The dispersion around zero provides an estimate of the measurement uncertainty on ${Z}^*_\text{T}(a_\text{T})$, which is typically below $0.1$~N$\cdot$m$^{-1}$. This level of precision is sufficient, as near-wall shear responses generally involve impedance values of a few N$\cdot$m$^{-1}$. Higher dispersion is observed at low amplitude, as the signal-to-noise ratio decreases (tangential-mode noise is typically of the order $\sim 10$ pm). Moreover, the small hysteresis observed between forward and backward sweeps originates from finite regulation times; reducing the sweep rate minimizes this effect at the expense of longer acquisition times.

\paragraph{Amplitude-dependent integration time:} Sweeping over three decades of amplitudes is more conveniently performed using logarithmic steps. However, the experimental software only allows linear stepping. To mitigate this limitation, the integration time is adapted depending on the amplitude: longer integration times are used at low amplitudes (up to $\tau_\text{int}=10$~s at $a_\text{T}\sim 20$~pm), while shorter integration times are used at high amplitudes (down to $\tau_\text{int}=20$~ms at $a_\text{T}\sim 20$~nm), with a typical step size of $\Delta a_\text{T}\sim 5$~pm. 

For improved visualization, a logarithmic binning procedure is applied \textit{a posteriori}, reducing typically 4000 data points acquired with linear spacing into about 50 averaged points on a logarithmic scale.

\subsection{Stiction offset}

Analysis of the amplitude sweeps revealed that introducing an additional fitting parameter in the form of a stiction offset $Z^{\prime\infty}_\mathrm{T}$ [N$\cdot$m$^{-1}$] can significantly improve the quality of the fits. In practice, this correction reduces the residuals and leads to more robust estimates of the viscoelastic parameters, particularly the relaxation timescale $\tau_0$.

The fitting function then becomes:

\begin{equation}
    Z^*_\text{T} = Z^0_\mathrm{T}\, \frac{(\omega_\text{T} \tau)^2 + j(\omega_\text{T} \tau)}{1+(\omega_\text{T} \tau)^2} + Z^{\prime\infty}_\mathrm{T}
    \label{eq:StictionOffset}
\end{equation}
where $Z^{\prime\infty}_\mathrm{T}$ is a purely real correction added to the Maxwell mechanical impedance. Physically, this contribution corresponds to an additional elastic stiffness that remains independent of the shear amplitude over the explored range.

The impact of this correction is amplified by the asymptotic behavior of the nonlinear viscoelastic model. At large amplitudes, the elastic component decreases as $Z^{\prime}_\text{T}\sim (a_\text{T})^{-2\nu}$, whereas the dissipative component follows the slower decay $Z^{\prime\prime}_\text{T}\sim (a_\text{T})^{-\nu}$. Consequently, even a small constant offset in stiffness can noticeably distort the high-amplitude scaling of $Z^{\prime}_\text{T}$ and bias the extracted value of $Z^0_\mathrm{T}$, while an equivalent offset in dissipation would have a much weaker influence on the fit quality.

\begin{figure}[htb!]
    \centering
    \includegraphics[width=17 cm, trim = 0cm 11cm 0cm 0cm, clip]{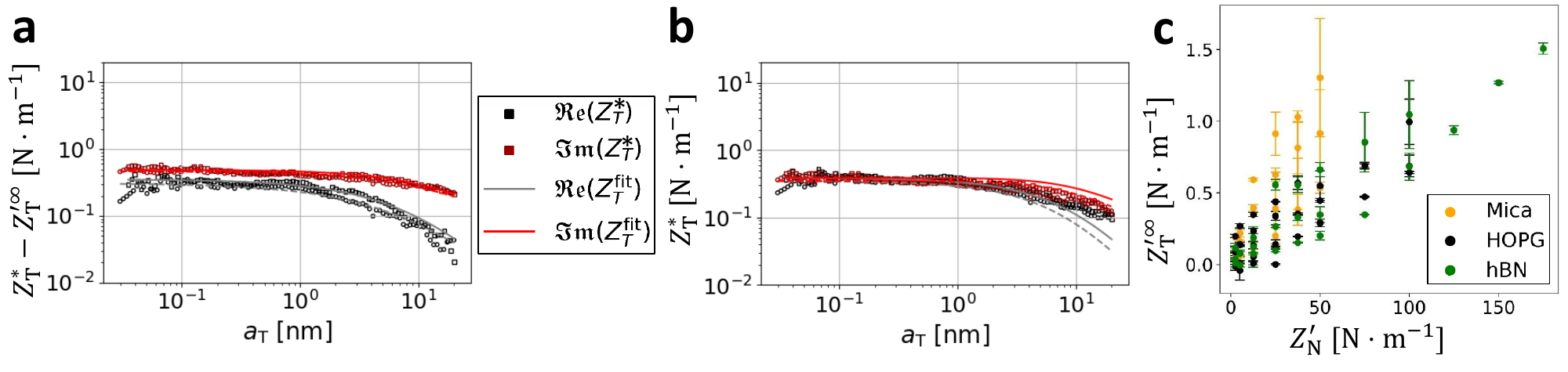}
    \caption{\textbf{Influence of a stiction offset.} Example of amplitude sweeps between tungsten and HOPG in pure water at load stiffness $Z^{\prime}_\text{N}=12.5$ N$\cdot$m$^{-1}$. \textbf{(a)} Including a stiction offset as a fitting parameter leads to $Z^{\prime\infty}_\mathrm{T}=0.071\pm 0.003$ N$\cdot$m$^{-1}$, then subtracting it to show $Z_\text{T}^*-Z^{\prime\infty}_\mathrm{T}$. The viscoelastic stiffness yields $Z^0_\mathrm{T}=1.03\pm 0.01$ N$\cdot$m$^{-1}$. \textbf{(b)} When no stiction offset $Z^{\prime\infty}_\mathrm{T}$ is included, or when it is forced to zero, the agreement with the nonlinear viscoelastic model is worse. The viscoelastic stiffness here yields $Z^0_\mathrm{T}=0.74\pm 0.04$ N$\cdot$m$^{-1}$. Uncertainties in values are the standard deviation between the fits of sweeps up and down. \textbf{(c)} Stiction offset $Z^{\prime\infty}_\mathrm{T}$ on mica, HOPG and hBN showing that $Z^{\prime\infty}_\mathrm{T}/Z^{\prime}_\text{N}\lesssim 2~\%$.}
    \label{fig:StictionOffset}
\end{figure}

Figure~\ref{fig:StictionOffset} illustrates the relevance of including a stiction offset $Z^{\prime\infty}_\mathrm{T}$ in the fitting procedure according to Eq.~\ref{eq:StictionOffset}. In Fig.~\ref{fig:StictionOffset}(a), subtracting a small stiction offset, approximately one order of magnitude smaller than the viscoelastic stiffness $Z^0_\mathrm{T}$, yields a much better agreement with the nonlinear viscoelastic model than in Fig.~\ref{fig:StictionOffset}(b), where $Z^{\prime\infty}_\mathrm{T}$ is constrained to zero. By more accurately reproducing the expected high-amplitude shear-thinning regime, the fit reveals a smaller Deborah number, $\text{De}=0.42\pm0.03$, whereas neglecting the offset leads to $\text{De}\sim1$ and a noticeably poorer fit quality. Accounting for this small correction therefore improves significantly the precision of the extracted relaxation timescale $\tau_0=\text{De}/\omega_\text{T}$ of confined water. The simultaneous fitting of the real and imaginary parts of the mechanical impedance thus represents a powerful approach and shows good agreement with the proposed model.

\paragraph{Resonance frequency shift.} Several mechanisms may contribute to such a stiction offset. First, a drift of the resonance frequency can be considered. This drift may originate from a slight displacement of the wetting line or from thermal fluctuations. In this case, verifying that $Z^{\prime}_\text{T}\approx0$ sufficiently far from contact, for instance at $D\sim1$~µm where no significant elastic interaction is expected, provides a practical way to identify and correct this systematic error.

\paragraph{Small tilt angle.} Second, the stiction offset could arise from elastic forces associated with the mechanical contact. If a small tilt angle $\theta$ exists between the tangential oscillation direction and the substrate plane, a contribution scaling as $Z^{\prime\infty}_\mathrm{T}\sim\tan(\theta)\,Z^{\prime}_\text{N}$ may be expected. In this framework, the stiction offset should increase proportionally to the normal load stiffness. For instance, assuming $|\theta|<10^\circ$ yields $Z^{\prime\infty}_\mathrm{T}/Z^{\prime}_\text{N}<0.18$. On Figure~\ref{fig:StictionOffset} (c), we can see that the ratio $Z^{\prime\infty}_\mathrm{T}/Z^{\prime}_\text{N}<2 \%$ on all three substrates which is consistent with a reasonably small tilt angle $\theta\lesssim 1^\circ$. Consequently, the measurement of a stiction offset increasing linearly with the load stiffness also confirms the rather good parallelism in the experimental setup.

\paragraph{Solid--solid contact lateral stiffness.} Third, the mechanical contact may induce elastic deformation of nanometric asperities present on the surface of the probe. Upon shear, these asperities could contribute an additional elastic stiffness scaling as

\begin{equation}
    Z^{\prime\infty}_\mathrm{T} \sim \frac{G^*A_\text{ssc}}{\delta_\text{ssc}}
\end{equation}

where $G^*$ [Pa] denotes the effective shear modulus, $A_\text{ssc}$ [m$^2$] the solid--solid contact area associated with asperities penetrating the water film, and $\delta_\text{ssc}$ [m] the local indentation depth at the scale of a contacting asperity. Since only a fraction of the nominal contact is expected to establish direct solid--solid contact, we assume $A_\text{ssc}<A$, where $A$ denotes the solid--liquid interaction area. For asperities of average radius $R_\text{asp}\ll R$, the contact area can be estimated as $A_\text{ssc}\sim\pi R_\text{asp}\delta$, where $\delta$ is the global indentation depth introduced in the soft-shell model. The local indentation depth is expected to be of the order of the RMS roughness, $\delta_\text{ssc}\sim h_\text{RMS}$. Combining these assumptions with the relationship $Z^{\prime}_\text{N}(\delta)$ from Eq.~\ref{eq:deGennes} yields:

\begin{equation}
    Z^{\prime\infty}_\mathrm{T}
    \sim
    \frac{G^*}{E^*}
    \frac{t}{h_\text{RMS}}
    \frac{R_\text{asp}}{R}
    Z^{\prime}_\text{N}
\end{equation}

This mechanism also predicts $Z^{\prime\infty}_\mathrm{T}\propto Z^{\prime}_\text{N}$, although the proportionality factor remains difficult to estimate accurately. As an order-of-magnitude estimate, taking $E^*\sim G^*$, $~t\sim10\,h_\text{RMS}$, and $R_\text{asp}\sim R/100$ leads to $Z^{\prime\infty}_\mathrm{T}/Z^{\prime}_\text{N}\sim0.1$, somewhat larger than the values typically observed experimentally.

Overall, introducing an additional fitting parameter in the form of a stiction offset improves the quality of several amplitude-sweep fits without compromising the validity of the nonlinear viscoelastic model or its applicability to the experimental data. Several mechanisms that could explain the origin of this correction have been considered, although the dominant contribution remains unclear. Consequently, a phenomenological interpretation of this parameter is adopted throughout the present work. Thereafter in the main text, in particular in Figure 2, we label $Z^*_\text{T}$ instead of $Z_\text{T}^*-Z^{\prime\infty}_\mathrm{T}$ for simplicity.

\newpage
\section{Contact mechanics and near-wall forces}
\subsection{van der Waals attraction (\textit{cf.} Fig. 1 d)}

Long-ranged attraction, characterized by a weak negative stiffness, can be observed for a separation distance $D > 3$ nm in Fig. 1 d. This attraction is attributed to van der Waals forces, scaling as:
\begin{equation}
Z^{\prime}_\text{N}(D) =- \frac{H R}{3 D^3}
\label{eq:vdW}
\end{equation}
where $H$ [J] is the Hamaker constant. Despite the low signal-to-noise ratio in the attractive regime, fitting (using Eq.~\ref{eq:vdW}) yields an estimate of $H = (5 \pm 1)\times10^{-20}$ J, which falls within the expected range of $10^{-21}$–$10^{-19}$ J reported in the literature  \cite{israelachvili2011intermolecular}. 

The mechanical zero for van der Waals attraction and solid contact estimation agree within an error margin smaller than 0.3 nm. This remaining uncertainty may originate from adsorbed water monolayers (estimated to a thickness of $\sim 0.25$ nm on each surface); or a subnanometric roughness of the tungsten oxide shell covering the probe, which effectively blurs the definition of the mechanical contact.

We highlight that the high rigidity of the tuning fork ($\sim40$ kN$\cdot$m$^{-1}$) allows us here to observe well-resolved attractive forces without the occurrence of any jump-to-contact instability which would be at play in more classical compliant systems.

Finally, we note that van der Waals attraction could not be systematically characterized, due to slight variation in probe roughness from one probe to another, blurring the faint attractive signal.

\subsection{Soft shell model (\textit{cf} Fig. 1 c)}

In Fig. 1 c, a linearly increasing stiffness can be observed upon indentation of depth $\delta\in[0,3]$ nm defined as $\delta=h_0-h>0$ in the contact region $h-h_0 <0$.

We note that in a sphere--plane geometry, classical Hertzian contact mechanics \cite{hertz1882ueber} would predict a normal repulsive force scaling as $F_\text{N} \propto \delta^{3/2}$, yielding a repulsive stiffness $Z^{\prime}_\text{N}=\partial F_\text{N}/\partial \delta\propto \sqrt{\delta}$. We thus tentatively interpret the linear scaling observed upon contact within a soft-shell contact model~\cite{de1996static} by assuming that the metallic indenter is covered by a thin, compliant or slightly porous oxide layer, leading to deviations from Hertzian behavior. The presence of such a softer layer would indeed lead to a quadratic force law, $F_\text{N} \propto \delta^2$, reflecting the modified mechanical response of the surface layer. 

We denote by $\delta$ [m] the indentation depth and by $R$ [m] the radius of curvature of the probe.  
The contact area $A$ [m$^2$] is written as $A=\pi a^2$, where the contact radius $a$ [m] scales as
\begin{equation}
    a=\sqrt{R\delta}
\end{equation}

Within the soft-shell framework, the pressure in the contact region is assumed to be homogeneous and is given by
\begin{equation}
    P(\delta)=\frac{E^*\,\delta}{2t}
\end{equation}
where $E^*$ and $t$ are respectively the reduced Young's modulus and thickness of the compliant layer, typically $t\sim 5$ nm (see Ref.~\cite{biegelsen1987ion}).  
Contrary to classical Hertzian contact mechanics, this homogeneous pressure distribution does not introduce geometric nonlinearities that would violate the Derjaguin approximation. The normal load force can therefore be readily computed as
\begin{equation}
    F_N(\delta)=\int_0^a P\,2\pi r\,\mathrm{d}r
    =\frac{\pi E^* R \delta^2}{2t}
\end{equation}

Because this force is conservative, it can equivalently be described by an effective spring of stiffness $Z^{\prime}_\text{N}$ [N$\cdot$m$^{-1}$], which can also be expressed as
\begin{equation}
    Z^{\prime}_\text{N} = 2\pi R\,P
\end{equation}
As a result, the load stiffness increases linearly with the indentation depth:
\begin{equation}
    Z^{\prime}_\text{N}(\delta)=\frac{\partial F_N}{\partial \delta}
    =\frac{\pi E^* R\,\delta}{t}
    \label{eq:deGennes}
\end{equation}

\begin{figure}[htb!]
    \centering
    \includegraphics[width=17cm, trim = 0cm 11cm 0cm 0cm, clip]{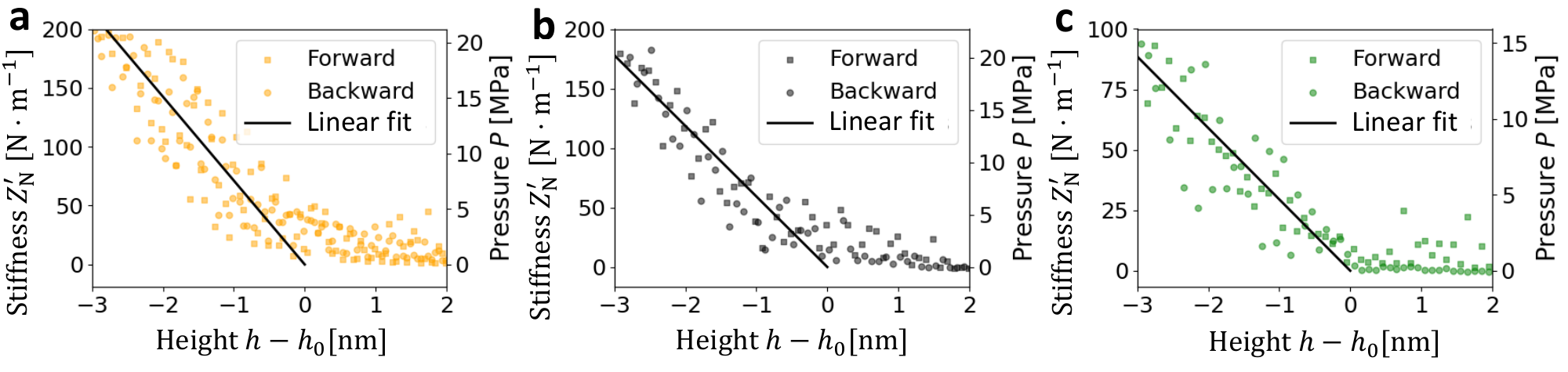}
    \caption{\textbf{Linear repulsion on three substrates} interpreted by the soft shell model:
    (a) mica yielding $E^*/t=15\pm1$ MPa/nm for $R\sim1.5$ µm;
    (b) HOPG yielding $E^*/t=13\pm1$ MPa/nm for $R\sim1.4$ µm;
    (c) hBN yielding $E^*/t=9\pm1$ MPa/nm for $R\sim1.0$ µm.
    Uncertainties correspond to the standard deviation between approach and retraction curves.}
    \label{fig:SoftShellRepulsion}
\end{figure}

As shown in Fig.~\ref{fig:SoftShellRepulsion}, a linear stiffness response is observed on all three substrates, with similar slopes in the range $E^*/t\sim 10$ MPa/nm. This supports the interpretation that the observed elastic behavior originates from a soft surface layer on the tungsten probe. Assuming an oxide thickness of $t = 5$ nm, the resulting effective elastic modulus is $E^* \sim 50$ MPa. This value is several orders of magnitude lower than the Young’s modulus of bulk tungsten ($\sim 400$ GPa), which can be attributed to the porous and compliant nature of the tungsten oxide layer. 

In practice, the geometric relation between indentation $\delta$ and contact radius $a$ allows extraction of the characteristic interaction area. The linear fit yields an estimation of the ratio $E^*/t$, from which the contact area can be directly estimated as
\begin{equation}
    A=\pi a^2=\frac{Z^{\prime}_\text{N}}{E^*/t}
\end{equation}

\subsection{Exponential repulsion (\textit{cf.} Fig. 1 e)}

In Fig. 1 e, the present work reports a near-wall monotonic, exponential-like repulsion, characterized by a nanometric decay length for a gap distance approximately $D\in [0, 3]$ nm. The load stiffness $Z^{\prime}_\text{N}$ [N$\cdot$m$^{-1}$] is fitted using:
\begin{equation}
    Z^{\prime}_\text{N}(D) = \frac{2\pi R W}{\lambda}\exp\!\left(-\frac{D}{\lambda}\right)
\end{equation}
where the separation is defined as $D = h-h_0$, with $h_0$ denoting the position of solid contact. Within the Derjaguin approximation ($D \ll R$), the surface energy $W$ [J$\cdot$m$^{-2}$] associated with this force corresponds to the maximal repulsion at $D=0$. This exponential-like repulsion can be ascribed to several effects, which we detail below.

\subsubsection{Hydration repulsion}
Structural forces, also called hydration forces in the case of water, are reported to exhibit monotonic exponential-like repulsion where the surface energy in the literature typically lies in the range $W \in [3,\,30]$ mJ$\cdot$m$^{-2}$ or below~\cite{israelachvili2011intermolecular}. The decay length $\lambda$ [m] is correlated with the molecular diameter and is therefore expected to fall within the range $\lambda \in [0.2,\,1]$ nm~\cite{ma2015origins,israelachvili2011intermolecular}.

\begin{figure}[h!]
    \centering
    \includegraphics[width=17cm, trim = 0cm 7.5cm 0cm 0cm, clip]{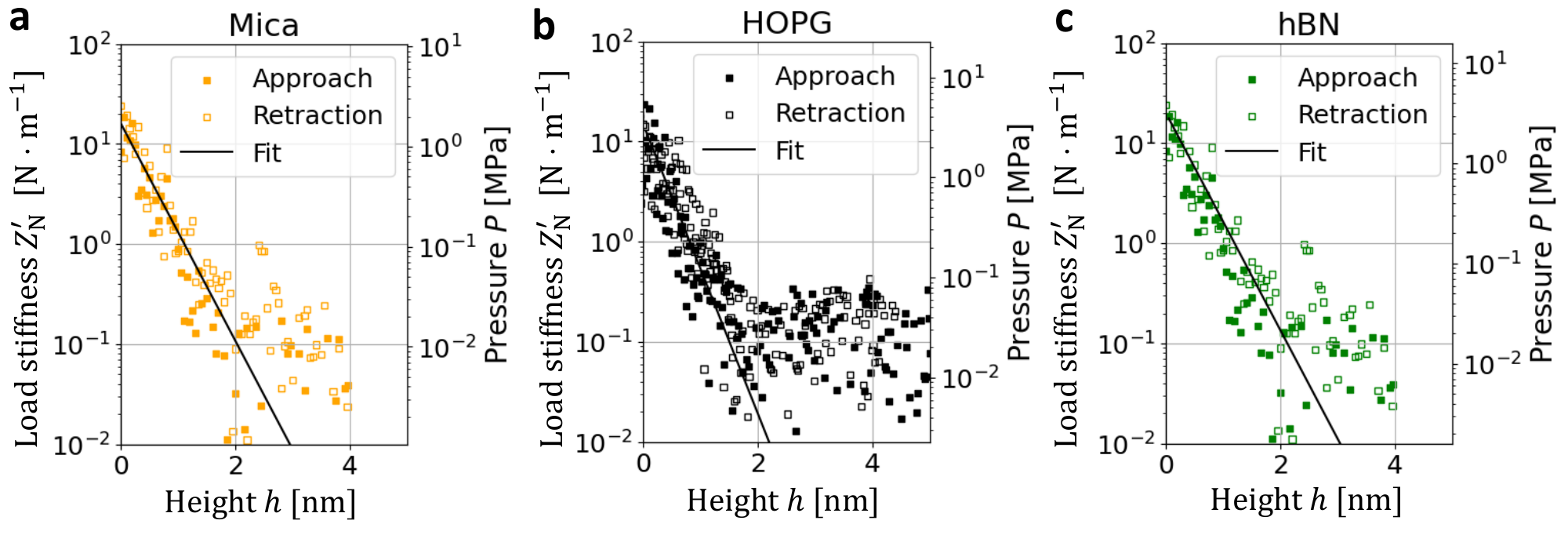}
    \caption{\textbf{Exponential repulsion} on three substrates:
    (a) mica yielding $\lambda = 0.4 \pm 0.1$ nm,
    (b) HOPG yielding $\lambda = 0.3 \pm 0.1$ nm,
    (c) hBN yielding $\lambda = 0.4 \pm 0.1$ nm.
    Uncertainties correspond to the standard deviation between forward and backward fits.}
    \label{fig:ExponentialRepulsions}
\end{figure}

Figure~\ref{fig:ExponentialRepulsions} shows good agreement between the experimental data and the exponential model. In all three cases, the estimated surface energy is $W \sim 1$ mJ$\cdot$m$^{-2}$, which may be underestimated due to the presence of a residual water film of thickness $e_0 = 0.5$ nm, corresponding to approximately two water monolayers. The decay length $\lambda\sim 0.4$ nm falls here within the expected range, with discrepancies of about 0.1~nm between independent experiments.

The present work suggests that exponential repulsion could be ascribed to hydration forces that arise as part of more general structural forces encountered for several liquid-solid interfaces. These structural forces can exhibit non monotonic profiles with oscillations characterizing the interlayer thicknesses\cite{greenwood2021effects}. However, the resolution of oscillatory solvation forces often requires ultra-sharp AFM tips\cite{siretanu2023ion}. Therefore, oscillatory forces may vanish on larger probes due to either surface roughness or to several solvation layers being probed at the same time around the apex of the probe, leaving only monotonic solvation repulsion\cite{israelachvili2011intermolecular}.

It is important to note that hydration forces are generally observed only in the presence of a sufficient electrolyte concentration. The present experiments were performed in ultrapure water, which was nonetheless partially carbonated through exposure to ambient air. In addition, the tungsten probe is covered by a native oxide layer, which is expected to undergo partial surface ionization. This surface charge likely promotes the adsorption and structuring of interfacial water molecules, thereby enhancing the hydration forces.

\subsubsection{Roughness-induced exponential repulsion}

Monotonic hydration forces are not the only possible explanation for the observed exponential repulsion. Indeed, when accounting for the probe’s roughness, an exponential-like repulsive behavior may also arise from a progressively increasing mechanical contact area.

According to Persson's theory of rough mechanical contacts~\cite{persson2007relation}, a randomly rough surface characterized by an RMS roughness (Sq) $h_\text{RMS}$ [m] and an RMS slope (Sdq) $h^{\prime}_\text{RMS}$ [-] can also generate an exponential repulsion at low pressure:

\begin{equation}
    P(D)\sim\epsilon E^*h^{\prime}_\text{RMS}\exp{\left(-\frac{D}{\gamma h_\text{RMS}}\right)}
\end{equation}

where $\epsilon$ and $\gamma$ are dimensionless constants of order unity. The characteristic decay length is therefore expected to scale as $\lambda \sim h_\text{RMS}$. Here, $D$ denotes the separation between the average spherical profile of the probe and the flat substrate. Because asperities can establish contact while the mean surfaces remain separated, mechanical interactions may already occur for $D>0$.

Within this framework, the apparent contact position $h_0$ corresponds to the crossover between the low-load regime described by Persson's theory and a higher-load regime where the density of contacting asperities progressively saturates, leading to a response more closely resembling the soft-shell model. Applying the Derjaguin approximation to the pressure law yields a corresponding load stiffness:

\begin{equation}
    Z^{\prime}_\text{N}(D)\sim 2\pi R \epsilon E^*h^{\prime}_\text{RMS}
    \exp{\left(-\frac{D}{\gamma h_\text{RMS}}\right)}
\end{equation}

The associated effective surface energy scales as

\begin{equation}
    W\sim \epsilon E^* h^{\prime}_\text{RMS}\gamma h_\text{RMS}
\end{equation}
while the decay length is given by $\lambda=\gamma h_\text{RMS}$. Assuming $\epsilon\sim\gamma\sim1$, the experimentally measured decay length $\lambda\sim0.4$ nm would correspond to a sub-nanometric RMS roughness of similar magnitude on the tungsten probe.

The prefactor further yields an estimate $E^*h^{\prime}_\text{RMS}\sim3$ MPa. At first sight, this value appears surprisingly small compared with the Young's modulus of tungsten ($\sim400$ GPa) or graphite ($\sim25$ GPa \cite{eskelsen2013correlating,yamanaka2000nanoscale}). It is however not that far from the effective modulus inferred from the soft-shell model, $E^*\sim50$ MPa.


Overall, Persson's framework provides a plausible alternative interpretation of the measured exponential repulsion. 
However, the present experiments do not allow a clear distinction between roughness-induced repulsion and hydration forces, since both mechanisms naturally produce exponential interactions with sub-nanometric decay lengths. Consequently, the analysis presented in this work remains essentially phenomenological, and both contributions may coexist in the measured force profiles.

\subsection{Neglected interactions}

The observed normal repulsion is attributed to soft shell repulsion, based on its linear profile, and characterizes the solid contact mechanics. Short range ($D\in [0,3]$ nm) exponential-like repulsion is also ascribed either to structural hydration forces or to low load Persson's contact mechanics. We discuss below the potential presence of other contributions in the normal force profile.

\paragraph{DLVO repulsion} First, DLVO repulsion arises from surface charge screening and also exhibits exponential decay for similarly charged substrates. However, its characteristic Debye length $\lambda_D$ depends strongly on ionic concentration, and for deionized water (18 M$\Omega\cdot$cm), potentially ionized by CO$_2$, $\lambda_D \sim 100$ nm, far exceeding the measured decay length $\lambda \lesssim 1$ nm. Regarding the amplitude of this DLVO repulsion, a typical uniform surface charge 
$\sigma \sim 1~\mathrm{mC \cdot m^{-2}}$ leads to a maximal stiffness $
Z^{\prime}_\mathrm{DLVO} \sim {4 \pi R \sigma^2}/{\varepsilon_0 \varepsilon} \sim 10^{-2}~\mathrm{N \cdot m^{-1}}$ for micrometric probes, which is buried in the noise due to insufficient sensitivity.

\paragraph{Elastohydrodynamics} Second, elastohydrodynamic (EHD) effects\cite{leroy2011hydrodynamic} may occur when the substrate deforms under high pressures in the confined liquid. If EHD dominated, dissipation would exceed the normal stiffness ($Z^{\prime\prime}_\text{N} > Z^{\prime}_\text{N}$), yet the data show $Z^{\prime}_\text{N} > Z^{\prime\prime}_\text{N}$ at all times, indicating that EHD is negligible compared to hydration repulsion and soft shell repulsion.

\paragraph{Normal viscoelasticity} Third, viscoelasticity could contribute in the normal direction, as it was observed tangentially. The normal oscillation amplitude $a_\text{N}=0.1 $ nm, is smaller than the size of a water molecule, so we may assume nonlinear effects are negligible. In a linear Maxwell viscoelasticity framework, the storage to loss ratio should scale as $Z^{\prime}_\text{N}/Z^{\prime\prime}_\text{N}\sim \omega_\text{N}\tau_0$. In the bulk, where relaxation timescales $\tau_0$ are very fast (in the picosecond range), the product $\omega_\text{N}\tau_0\ll1$, leading to a regime where normal dissipation $Z^{\prime\prime}_\text{N}$ largely exceeds the normal stiffness $Z^{\prime}_\text{N}$. Although it could slightly contribute, because $Z^{\prime}_\text{N} > Z^{\prime\prime}_\text{N}$ was systematically observed, normal viscoelasticity was consequently neglected.


\newpage
\section{Additional analysis}

\begin{figure}[htb!]
    \centering
    \includegraphics[width=12 cm, trim = 0cm 0cm 15cm 0cm, clip]{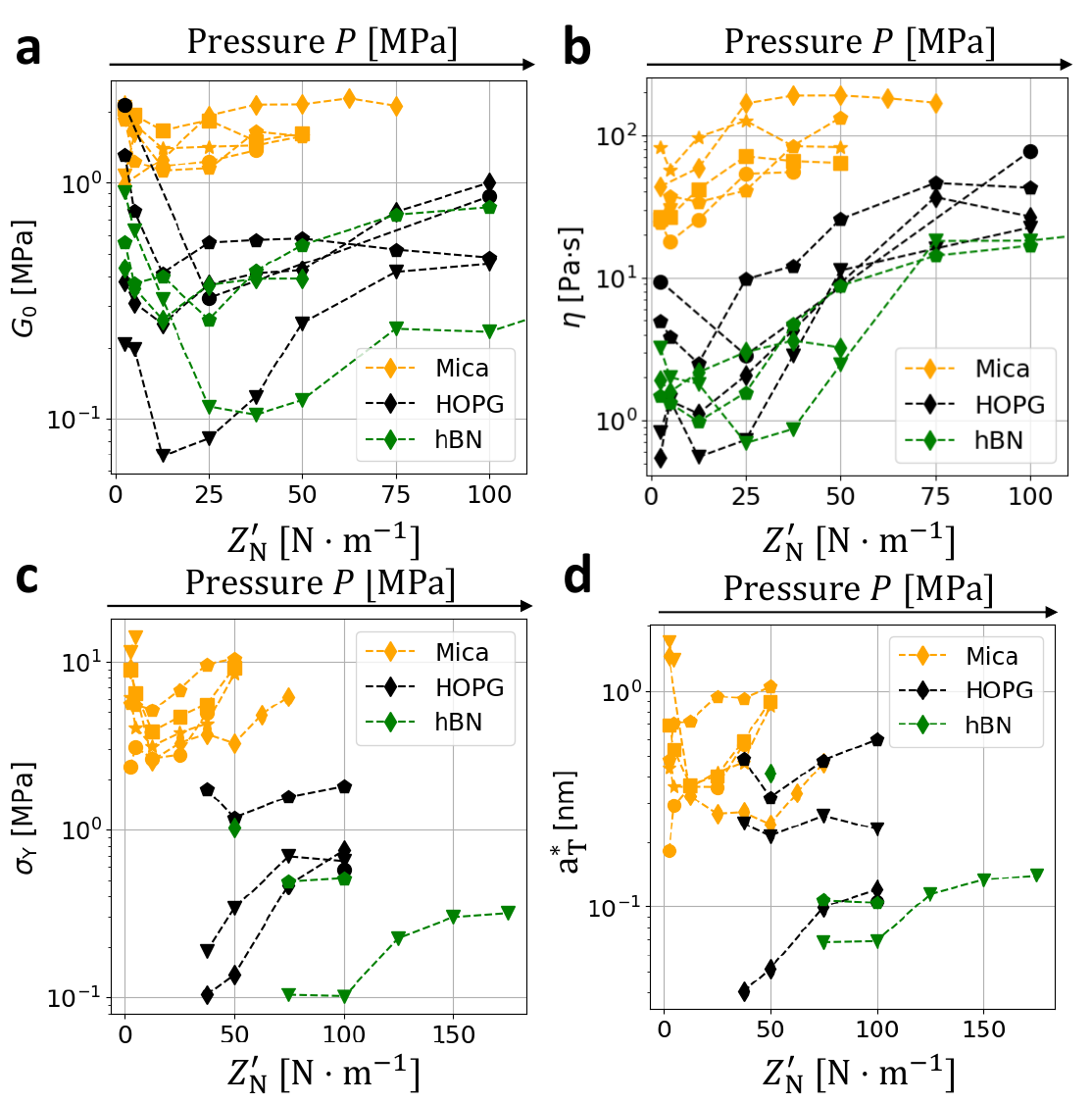}
    \caption{Estimation of nonlinear viscoelastic properties of nanoconfined water from fitting parameters:
    (a) \textbf{Viscoelastic modulus $G_0$ [Pa]},
    (b) \textbf{Structural viscosity $\eta$ [Pa$\cdot$s]}, (c) \textbf{Yield stress} and (d) \textbf{Yielding shear amplitude}
    }
    \label{fig:FurtherAnalysis}
\end{figure}

\subsection{Viscoelastic modulus $G_0$ [Pa]}

The viscoelastic modulus $G_0$, presented in an inset of Fig 3 b, is estimated from the viscoelastic stiffness $Z^0_\mathrm{T}$ and geometric factors according to the expression:

\begin{equation}
    G_0 = \frac{Z^0_\mathrm{T}\,e_0}{A}
\end{equation}
where $e_0$ is the thickness of the adsorbed water, taken as 0.5 nm, and $A$ is the interaction area inferred from contact mechanics.

Figure~\ref{fig:FurtherAnalysis}(a) shows the estimated viscoelastic modulus $G_0$, which is relatively insensitive to load stiffness and significantly higher on mica than on hBN and HOPG. Its magnitude lies in the MPa range, indicating a response much softer than bulk water, as well as far below the Young’s modulus of crystalline substrates or tungsten.

\subsection{Structural viscosity $\eta$ [Pa$\cdot$s]}

Even though the present work describes the highly non-Newtonian behavior of confined interfacial water, the energy dissipation can be accounted for in a more general sense through a structural viscosity $\eta$~[Pa$\cdot$s] describing the dissipation of the fluid at vanishing shear. In viscoelastic materials, the relaxation time $\tau_0$ scales with the structural viscosity as $\tau_0 \sim \eta/G_0$, which reveals a dramatic increase of the local viscosity of interfacial water compared to bulk water on all three investigated substrates. Using the relation $\eta = \tau_0 G_0$, the estimated viscosity of interfacial water on yield values of order $\eta \sim 1$--$100$~Pa$\cdot$s, corresponding to an enhancement of $10^{3}$--$10^{5}$ relative to bulk water. Figure~\ref{fig:FurtherAnalysis}(b) shows the evolution of the estimated structural viscosity as a function of the applied load, where $\eta$ on mica appears typically one order of magnitude higher than on HOPG and hBN. The progressive increase of the structural viscosity with load or pressure is consistent with a progressive slowdown of water upon molecular confinement.

Interestingly, if the viscoelastic modulus had remained in the bulk-water range ($G_0 \sim 1$~GPa), the inferred viscosity would instead be consistent with the results of Cafolla \textit{et al.}\cite{cafolla2018lubricating}, who reported that the local viscosity of pure water on mica reaches a maximum of $\sim 2 \times 10^{7}\,\eta_\mathrm{bulk}$ at a confinement gap of $D \sim 0.5$~nm, corresponding to approximately two water monolayers. However, using the experimentally extracted modulus $G_0 \sim 1$~MPa in the present work yields a maximum viscosity of $\eta \sim 10^{2}$~Pa$\cdot$s, which is closer to the values reported by Li \textit{et al.}\cite{li2007structured}.

\subsection{Yield stress $\sigma_\text{Y}$ [Pa]}
The yield stress $\sigma_\text{Y}$ is computed for yield-stress regimes according to the expression:
\begin{equation}
    \sigma_\text{Y} = \frac{Z^{\prime}_0\,a_\text{T}^*}{A}
\end{equation}
where $a_\text{T}^*$ [m] is the yielding amplitude and $Z^{\prime}_0$ is the low-amplitude stiction representing the amount of elastic energy stored in the solid-like state. For high Deborah number ($\text{De}>1$), we assimilate $Z^{\prime}_0\simeq Z^0_\mathrm{T}$, the viscoelastic modulus estimated as a fitting parameter. We define then the product $F_\text{Y}=Z^0_\mathrm{T}\,a^*_\text{T}$ as the yielding force.

As shown in Fig.~\ref{fig:FurtherAnalysis}(c), the corresponding yield stresses, $\sigma_\mathrm{Y} = F_\mathrm{Y}/A$, remain relatively low despite the highly confined state of the water film. Typical values lie in the range
$\sim 10^6$--$10^7$~Pa on mica and $\sim 10^5$--$10^6$~Pa on HOPG and hBN. The systematically lower yield stresses measured on HOPG and hBN are consistent with the enhanced lubricity of hydrophobic substrates and suggest that water confined at these interfaces can be fluidized more easily than on hydrophilic mica.

A useful perspective is to compare these values with the yield stresses of other soft glassy materials. The measured stresses are several orders of magnitude larger than those typically reported for colloidal suspensions, foams, or emulsions ($10^0$--$10^3$~Pa), yet remain well below those of crystalline solids. This intermediate behavior further supports the view that nanoconfined water behaves as a soft solid or yield-stress fluid, rather than as a conventional liquid.

We now consider a possible microscopic
interpretation based on a stress-assisted, thermally activated energy-barrier mechanism. Within this framework, water molecules are trapped in a corrugated energy landscape arising from both intermolecular interactions and interactions with the confining walls. Flow occurs only when the applied shear stress is sufficient to lower the effective barrier and activate molecular rearrangements. Increasing the normal load increases the local pressure and strengthens the confinement-induced ordering of the water film, thereby increasing the corrugation potential and the corresponding yield stress. The lower yield stresses observed on HOPG and hBN may then reflect a smoother effective energy landscape, consistent with the low friction evidenced on these hydrophobic interfaces.

\subsection{Yielding shear amplitude $a_\text{T}^*$ [m]}

As shown in Fig.~\ref{fig:FurtherAnalysis}(d), the yielding transition generally occurs at a critical amplitude $a_\text{T}^*$ comparable to or larger than the diameter of a water molecule ($\simeq 0.25$ nm), particularly on mica. In contrast, on HOPG and hBN, fluidization is sometimes observed at substantially smaller amplitudes, with values reaching only a few tens of picometers (see main text for discussion).

\newpage
\section{Logarithmic friction}

The present results lie at the intersection of nanorheology and nanotribology. While the shear response is quantitatively described here using a nonlinear viscoelastic model, consistent with a nanorheological framework, the large-amplitude behavior may also be interpreted from a nanotribological perspective. Indeed, at high oscillation amplitudes, the loss impedance typically scales as $Z^{\prime\prime}_\text{T}\sim a_\text{T}^{-\nu}$. Since $\nu \lesssim 1$, the dissipative force $F_\text{T}^{\text{D}} = Z^{\prime\prime}_\text{T} \cdot a_\text{T}$ depends only weakly on the oscillation amplitude, yielding $F_\text{T}^{\text{D}}\sim a_\text{T}^{\,1-\nu}$ in the high amplitude regime. This scaling is therefore not substantially different from the logarithmic velocity-dependence $F_\text{T}^{\text{D}}\sim \ln(v_\text{T})$ with $v_\text{T}\sim a_\text{T}\omega_\text{T}$ frequently reported in nanotribology experiments on confined water \cite{diao2019slippery,ma2015origins} at constant velocity $v_\text{T}$. Such logarithmic friction laws have been interpreted in terms of shear-assisted, thermally activated barrier crossing within an Eyring framework \cite{spikes2018stress,gnecco2000velocity}.

\begin{figure}[h!]
    \centering
    \includegraphics[width=17cm, trim = 0cm 8cm 3cm 0cm, clip]{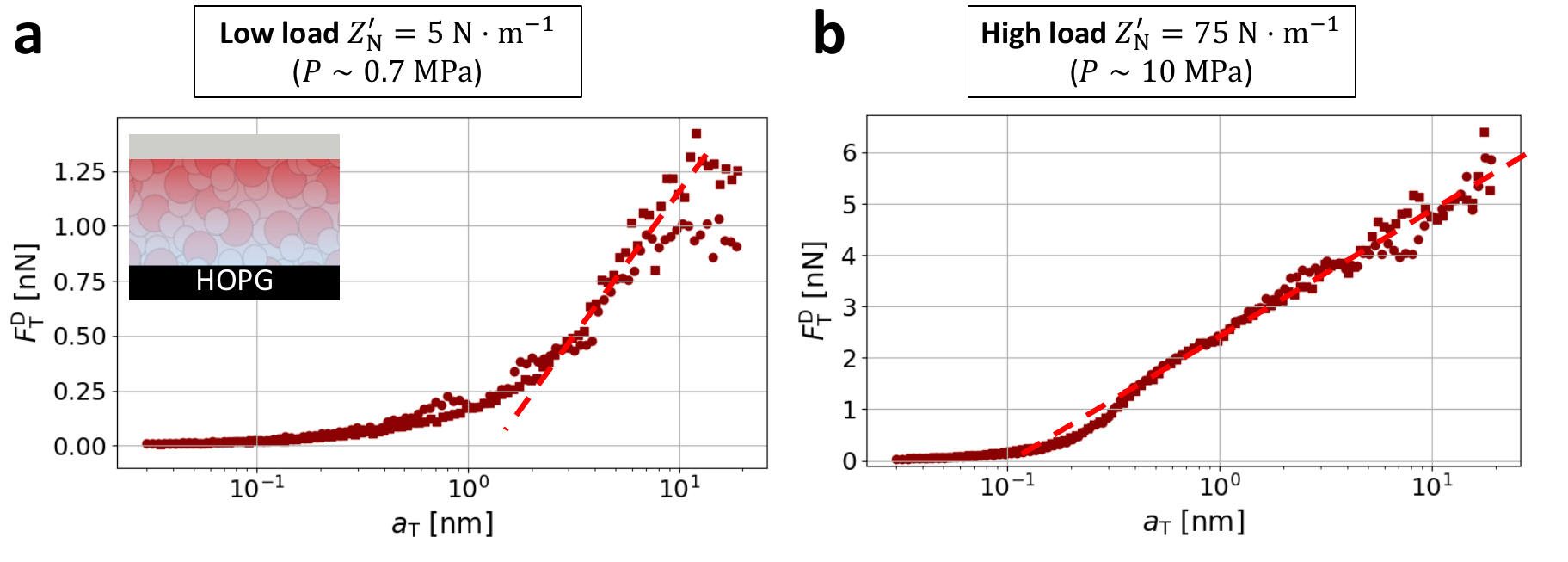}
    \caption{\textbf{Logarithmic friction force} $F^\text{D}_\text{T}$ measured on HOPG in pure nanoconfined water. Red circles: increasing $a_\text{T}$. Red squares: decreasing $a_\text{T}$. Red dashed line: guide for the eye on the locally logarithmic friction for $a_\text{T}>a_c$, with $a_c$ defined as the amplitude where a strong deviation from logarithmic friction appears. (a) Low-load regime, $Z^{\prime}_\text{N}=5$~N$\cdot$m$^{-1}$, corresponding to a pressure of approximately $P\sim0.7$~MPa, with $R\simeq 1.2$~µm. (b) High-load regime, $Z^{\prime}_\text{N}=75$~N$\cdot$m$^{-1}$, corresponding to $P\sim10$~MPa. Black dashed line: the yielding amplitude $a_\text{T}^*$, below (above) which lies a solid-like (liquid-like) state, coincides approximately with the critical amplitude $a_c$.}
    \label{fig:LogFriction}
\end{figure}

To assess the relevance of this model, we compare its predictions with experimental measurements obtained on HOPG at different normal loads, as shown in Figure~\ref{fig:LogFriction}. 

Figure~\ref{fig:LogFriction}(a) first illustrates the low-load regime, previously identified as exhibiting shear-thinning rheology. In this case, the experimental data are poorly described by a logarithmic friction, except at amplitudes larger than a critical amplitude $a_c\sim$ 2 nm. At lower shear amplitude $a_\text{T}<a_c$, the response is better described by a viscous friction. Consistently, the nonlinear viscoelastic analysis yields a Deborah number $\mathrm{De}=\omega_\text{T}\tau_0<1$, indicating that relaxation occurs on a timescale shorter than the oscillation period. Interestingly, the characteristic amplitude extracted from the tribological model is found to be comparable to the critical amplitude obtained from the rheological analysis, i.e., $a_c \simeq a_\text{T}^*$, in the nanometer range. 

In contrast, for the high-load regime shown in Figure~\ref{fig:LogFriction}(b), a clearer logarithmic friction  can be observed for $a_\text{T}>a_c\simeq a_\text{T}^*$. This range corresponds to the liquid-like branch of the yield-stress regime associated with this curve. However, in the solid-like state at shear amplitude lower than the yielding amplitude, the dissipation deviates strongly from logarithmic friction.

In summary, Eyring mechanisms may provide the basis for a theoretical framework capable of describing the soft-glassy rheology of nanoconfined water. The model leading to a logarithmic friction shows good agreement with the experimental data at high enough shear amplitude, especially at high loads. However the significant deviation from logarithmic dependency at lower amplitude shows the limitation of the model in our experiments.  Extending this framework to account for the whole shear response therefore remains an important direction for future work.





\clearpage
\section{Supplementary References}

\bibliography{SuppRef}